\documentclass[a4paper,11pt]{article}
\usepackage[utf8x]{inputenc}
\usepackage[margin=0.9in]{geometry}
\usepackage{amsmath}
\usepackage{latexsym}
\usepackage{amssymb}
\usepackage[english]{babel}
\usepackage{color}
\usepackage[multiple]{footmisc}
\usepackage{jheppub}
\pdfoutput=1
\usepackage{jheppub}
\usepackage{breakurl}
\usepackage{mathrsfs}
\usepackage{amsmath}
\usepackage{amsfonts}
\usepackage{mathrsfs}
\usepackage{slashed}
\usepackage{epigraph}
\usepackage{amssymb}% http://ctan.org/pkg/amssymb
\usepackage{pifont}% http://ctan.org/pkg/pifont
\usepackage{fancyhdr}
\usepackage{amssymb}
\usepackage{graphicx}
\usepackage{longtable}
\usepackage{makecell}
\usepackage{amscd}
\usepackage{tabu}
\usepackage{bm}
\usepackage{enumerate}
\usepackage{caption}
\usepackage{subcaption}
\usepackage{bbold}
\usepackage{hyperref} 
\hypersetup{linktocpage=true}
\usepackage[all]{hypcap}
\hypersetup{
   bookmarks=true,         % show bookmarks bar?
   unicode=false,          % non-Latin characters in Acrobat bookmarks
   pdftoolbar=true,        % show AcrobatÃs toolbar?
   pdfmenubar=true,        % show AcrobatÃs menu?
   pdffitwindow=false,     % window fit to page when opened
   pdfstartview={FitH},    % fits the width of the page to the window
   pdftitle={My title},    % title
   pdfauthor={Author},     % author
   pdfsubject={Subject},   % subject of the document
   pdfcreator={Creator},   % creator of the document
   pdfproducer={Producer}, % producer of the document
   pdfkeywords={keyword1} {key2} {key3}, % list of keywords
   pdfnewwindow=true,      % links in new window
   colorlinks=true,        % false: boxed links; true: colored links
   linkcolor=blue,         % color of internal links
   citecolor=magenta,      % color of links to bibliography
   filecolor=magenta,      % color of file links
   urlcolor=cyan           % color of external links
}

\usepackage{fancyhdr}
\usepackage{booktabs}
\usepackage{tikz}
\usetikzlibrary{decorations}
\usetikzlibrary{arrows,intersections,shapes,decorations.markings}
\usetikzlibrary{arrows.meta}
\usetikzlibrary{shapes.misc}
\usetikzlibrary{decorations.pathmorphing}

\preprint{MITP/18-019
}
\title{A flavoured dark sector}
\author{Sophie Renner}
\author{and Pedro Schwaller}

\affiliation{\normalfont{PRISMA Cluster of Excellence \& Mainz Institute for Theoretical Physics, Johannes Gutenberg University, 55099 Mainz, Germany}}
\emailAdd{sorenner@uni-mainz.de}
\emailAdd{pedro.schwaller@uni-mainz.de}

\abstract{We explore the phenomenology of a QCD-like dark sector which confines around the GeV scale. 
The dark sector inherits a flavour structure from a coupling between dark quarks and SM quarks via a heavy mediator, which leads to exciting new phenomena. 
While stable baryonic bound states are the dark matter candidates, the
phenomenology is dominated by the lightest composite mesons, the dark pions, which can have decay lengths ranging from millimetres to hundreds of meters. For masses below 1.5~GeV, their exclusive decays to SM mesons are calculated for the first time by matching both dark and visible sectors to a chiral Lagrangian. 
Constraints from big bang nucleosynthesis, dark matter direct detection and flavour single out a small region of allowed parameter space for dark pion masses below 5~GeV. It is best probed by the fixed target experiments NA62 and SHiP, where dark pions can be produced copiously in rare decays like $B\to K \pi_D$. 
The dominant $\pi_D \to K^\pm \pi^\mp$ and $\pi_D \to 3 \pi$ decay modes are a smoking gun for a CP-odd, flavour violating new resonance. 
Heavier dark pions are best searched for at the LHC, where they decay after hadronisation to produce jets which emerge into SM states within the detector. Here the flavour structure ensures different flavours emerge on different length scales, leading to a striking new feature in the emerging jets signature. 
}

\begin{document}

\maketitle

\section{Introduction}

The origin and nature of dark matter is one of the biggest open questions of contemporary particle physics. Given the complexity of the Standard Model (SM) -- the visible sector of the universe -- it would not be surprising if the dark matter is also just one component of a larger \textit{dark sector}. Indeed many extensions of the SM feature dark sectors with new forces and symmetries. 

Here we are concerned with a QCD-like dark sector, i.e.~a new non-abelian $SU(N)$ symmetry with $n_d$ dark quarks in the fundamental representation and a confinement scale $\Lambda_D$ around the GeV scale. Such hidden sectors appear for example in the context of twin Higgs models~\cite{Chacko:2005pe,Craig:2015pha,Barbieri:2015lqa}, composite and/or asymmetric dark matter~\cite{Blennow:2010qp,Frandsen:2011kt,Buckley:2012ky,Hambye:2013sna,
Bai:2013xga,Cline:2013zca,Boddy:2014yra,Hochberg:2014kqa,Antipin:2015xia,Garcia:2015loa,Hochberg:2015vrg,Dienes:2016vei,Lonsdale:2017mzg,Davoudiasl:2017zws,Berlin:2018tvf}, and are also ubiquitous in string theory~\cite{Halverson:2016nfq,Acharya:2017szw}. 

As was already realised in the seminal works~\cite{Strassler:2006im,Strassler:2006ri}, the phenomenology of such models depends crucially on how they are coupled to the visible sector, i.e.~on the so called mediators. Besides neutral mediators that couple to SM singlet operators\footnote{The gauge invariant SM operators of dimension $<4$ are the Higgs portal $|H|^2$, the neutrino portal $LH$ and the kinetic mixing portal $F_{\mu\nu}$ (or $B_{\mu\nu}$) to which one can renormalizably couple a new scalar, fermion or vector boson, respectively~\cite{Alexander:2016aln}. 
}, new particles charged under the SM interactions can connect the visible and dark sectors.

In this work, we study for the first time the flavour structure that is imposed on a non-abelian dark sector by a bi-fundamental scalar mediator which is charged under both QCD and the dark $SU(N)$ symmetry. This type of mediator was introduced in~\cite{Bai:2013xga} and shown to lead to a new collider signature called \textit{emerging jets} in~\cite{Schwaller:2015gea}, however the flavour structure of the coupling to the dark sector was neglected in those studies. Here we show that it has quite dramatic consequences for the phenomenology of these models, but also provides new ways of searching for them. 

After introducing the model in Sec.~\ref{sec:model}, we compute lifetimes and branching ratios of dark pions in the flavoured case in Sec.~\ref{sec:darkmeson}, including some subtleties regarding the decay of GeV and lighter states into SM hadrons which requires using a chiral Lagrangian for both the dark and visible sector. We then identify the regions of parameter space consistent with $\Delta F = 2$ flavour violating processes and impose constraints arising from $\Delta F = 1$ flavour violating $B$ and $K$ meson decays (Sec.~\ref{sec:flavour}) as well as from cosmology (Sec.~\ref{sec:cosmology}). In Sec.~\ref{sec:emerging} we discuss the impact of our results on collider searches for dark QCD models. New ways for probing this model at fixed target experiments are then proposed in Sec.~\ref{sec:fixedtarget}.

Before continuing with the description of the model, a comment regarding the flavoured dark matter paradigm~\cite{Batell:2011tc,Agrawal:2011ze,Calibbi:2015sfa} is in order. Our model has some similarities with the flavoured dark matter model discussed in~\cite{Agrawal:2014aoa}, however there is also a crucial difference. In our model, dark matter stability is guaranteed by a conserved $U(1)$ dark baryon number, i.e.~flavour is not necessary for dark matter to survive. Instead the interesting flavoured phenomenology arises from the behaviour of the composite dark pions which are not protected by dark baryon number (since they are particle-antiparticle bound states), and which transform in the adjoint of the dark flavour symmetry. As a consequence, while the constraints arising from $\Delta F = 2$ constraints are somewhat similar in our model, the $\Delta F = 1$ constraints and phenomenological signatures are vastly different.

\section{The model}
\label{sec:model}

\begin{figure}
\centering
\includegraphics[trim = 10 10 10 60,clip,width=0.7\textwidth]{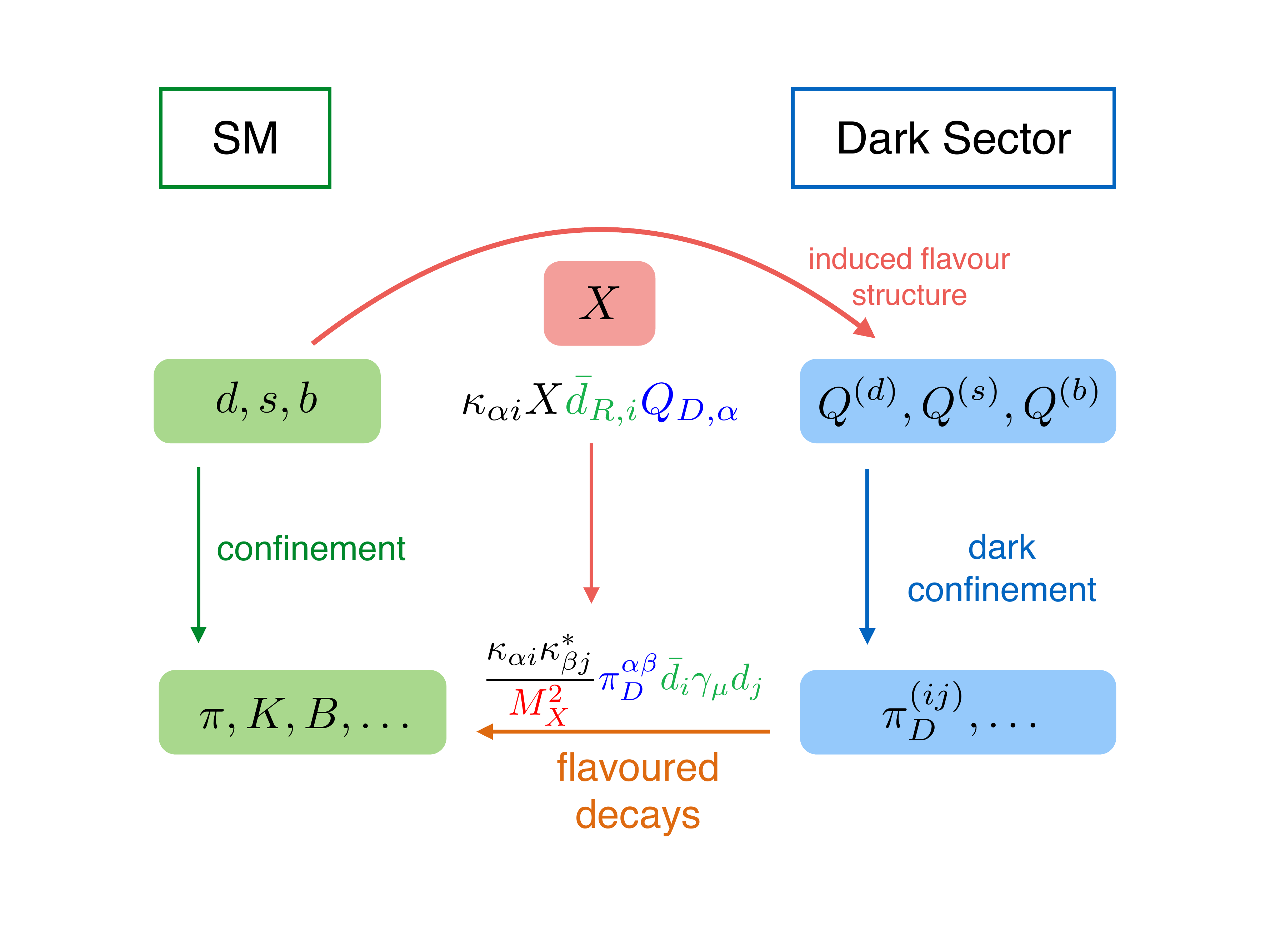}
\caption{The composite dark sector communicates with the SM through a flavoured dark portal: the mediator $X$ is a bi-fundamental scalar coupling to quarks and dark quarks, thereby inducing a flavour structure in the dark sector. The main consequence is that the dark pions, which are the lightest composite dark sector states, have lifetimes and branching ratios to SM particles which now depend on their flavour composition. 
\label{fig:model}}
\end{figure}

The gauge group of the SM ($G_{\rm SM}$) is extended to
\begin{align}
	G_{\rm SM} \times SU(N_d)\,,
\end{align}
where $N_d$ is the number of dark colours. We further introduce $n_d$ dark quarks $Q$, which are singlets under $G_{\rm SM}$ and transform in the fundamental of $SU(N_d)$. 
%
%Our model consists of a dark sector with a non-abelian $ SU(N_d)$ gauge symmetry and $n_d$ Dirac fermions $Q$ transforming in the fundamental representation of $ SU(N_d)$ and which are neutral under the SM gauge groups, which we will also refer to as dark quarks. 
For $n_d \leq 4 N_d$ the theory confines at a scale $\Lambda_D$. For all practical purposes we will set $N_d = 3$ in the following. The Lagrangian of the dark sector takes the form
\begin{align}
	{\cal L} = - \frac{1}{4} (G_D^{\mu\nu,a})^2 + \bar{Q}_\alpha i {D\!\!\!\!/\,\,} Q_\alpha - m_{Q,\alpha\beta} \bar{Q}_\alpha Q_\beta\,,
\end{align}
where $G_D$ is the dark gluon field strength tensor, and $\alpha, \beta$ are dark flavour indices. 

We are mostly interested in the case where the dark flavour symmetry is only weakly broken by the dark quark mass term $m_{\alpha\beta}$ ($m_{\alpha\beta} \ll \Lambda_D$), such that the lightest states in the dark sector are $n_d^2-1$ Goldstone bosons - dark pions - with masses $m_{\pi_D} \ll \Lambda_D$, which arise from the breaking of the $SU(n_d) \times SU(n_d)$ chiral symmetry by the dark QCD condensate. 

Communication between the dark and visible sectors is established through a bi-fundamental scalar field $X$ which transforms as $(3,N_d)$ under $SU(3)_{\rm colour} \times SU(N_d)$. Such bi-fundamentals are required e.g.~in the dark QCD model~\cite{Bai:2013xga} and could easily appear in UV completions of twin Higgs models~\cite{Barbieri:2015lqa,Geller:2014kta} or models where the dark gauge symmetry unifies with QCD at some higher scale. Collider constraints require that the mediator mass $M_X \gtrsim $~TeV, while the confining dark sector, being SM neutral, can be significantly lighter. We will in particular identify the viable parameter space for $\Lambda_D$ below the weak scale; a light dark sector. The structure of the model at high and low scales is displayed in Figure~\ref{fig:model}. As can be seen, both sectors undergo confinement, and can be treated using chiral perturbation theory for both visible and dark sectors. 

From a bottom up perspective, it is also useful to think of $X$ as a $t$-channel alternative to the usually considered $Z'$ or Higgs portal mediators to a dark sector. As we will discuss in more detail below, this has dramatic consequences for the properties of the dark sector bound states. In particular, if the quantum numbers of $X$ are such that Yukawa couplings of the form
\begin{align}
	{\cal L}_{\rm yuk} \supset - \kappa_{\alpha i} \bar{q}_i Q_{\alpha} X
\end{align}
are allowed, with $q$ any SM quark field, this imposes a flavour structure on the interactions of the dark quarks. In~\cite{Schwaller:2015gea} the coupling to right-handed down type quarks was considered, and  $\kappa_{\alpha i} \sim {\cal O}(1)$ was assumed such that all dark mesons ended up having the same lifetime, but neglecting possible constraints on $\kappa_{\alpha i}$ from flavour physics. The main goal of this section is to investigate the flavour structure of $\kappa_{\alpha i}$.

With hypercharge $Y_X = 1/3$, the only possible Yukawa coupling has the form
\begin{align}
\mathcal{L} \supset - \kappa_{\alpha i} \bar{d}_{Ri} Q_{L\alpha} X + h.c.\,,
\end{align}
which explicitly breaks down-quark as well as dark-quark flavour symmetries. Alternatively, a Yukawa coupling to up-type quarks or to left-handed quark doublets are also possible by choosing $Y_X = -2/3$ and $Y_X = -1/6$, respectively. In this work we make the choice $Y_X = 1/3$ for phenomenological reasons; coupling to right-handed down quarks allows for interesting effects in flavour observables, for example in $B$ decays, while keeping the flavour structure as simple as possible.  
In the absence of dark gauge interactions, similar couplings were studied before in the context of flavoured DM~\cite{Agrawal:2014aoa,Jubb:2017rhm,Blanke:2017fum}.

Using singular value decomposition, the matrix $\kappa$ can be written as
\begin{align}
	\kappa=VDU\,,
\end{align}
where $U$ is a $3 \times 3$ unitary matrix, $V$ is a $n_d \times n_d$ unitary matrix and $D$ is a $n_d\times 3$ non-negative diagonal matrix. If all the dark quarks have the same Lagrangian mass term, i.e.~$m_{Q,\alpha\beta} = m_Q \delta_{\alpha\beta}$ in some basis, there is a  $U(n_d)_{\rm dark}$ flavour symmetry in the dark sector, unbroken by any pure-dark Lagrangian terms, which can be used to rotate $V$ away. In the following we will assume that this is the case, meaning that the Yukawa couplings $\kappa$ are the only source of dark flavour symmetry breaking.

An immediate consequence of this is the following: If $n_d > 3$, there is an unbroken $U(n_d - 3)$ symmetry in the dark sector, which makes one or more dark pions stable.\footnote{While this symmetry may be broken by the WZW term, at the lowest order it mediates interactions between at least five dark pions (since photons don't couple to dark quarks), so the least suppressed decay mode of the stable dark pions will be to eight SM quarks, suppressed by $M_X^{-16}$.} Therefore, in the following we will restrict ourselves to the case of $n_d =3$, and leave the case of flavour stabilised dark pion dark matter for a future study. 

The matrix $U$ can be further decomposed into three unitary rotation matrices 
\begin{equation}
U=U_{23}U_{13}U_{12}\,,
\end{equation}
where $U_{ij}$ is the matrix that rotates $i \leftrightarrow j$, for example
\begin{equation}
U_{12} = \left(\begin{matrix}
c_{12} & s_{12}e^{-i \delta_{12}} & 0 \\
- s_{12}e^{-i \delta_{12}} & c_{12} & 0 \\
0 & 0 & 1
\end{matrix} \right),
\end{equation}
introducing the mixing angles $\theta_{ij}$ via $s_{ij}= \sin \theta_{ij}$, $c_{ij}= \cos \theta_{ij}$ and CP phases $\delta_{ij}$.  
Furthermore it is convenient to parameterise the diagonal matrix $D$ as follows~\cite{Agrawal:2014aoa}:
\begin{equation}
D=\bigg(\kappa_0\cdot \mathbb{1}+ \textrm{diag}(\kappa_1,\kappa_2,-(\kappa_1+\kappa_2))\bigg).
\end{equation}
The non-negativity of $D$ implies $\kappa_0 \geq 0$ and $|\kappa_1 + \kappa_2| \leq \kappa_0$. 
In the limit where $D$ is proportional to the identity matrix, $U$ and $D$ commute and therefore $\kappa \propto \mathbb{1}$ by choosing $V = U^\dagger$. In other words in this case a full $SU(3)$ subgroup of the $SU(3)_d \times SU(3)_{\rm dark}$ flavour symmetry remains unbroken. We will refer to this scenario as the \textit{alignment limit}. 

If, instead of transforming under their own flavour symmetry group, the dark quarks were assigned to representations of the SM flavour group, the alignment limit would correspond to a minimally flavour violating (MFV) scenario in which the charges of the dark quarks and the $\kappa$ matrix are chosen to be
\begin{align}
Q \sim (1,1,3)\,, && \kappa \sim (1,1,1)\,,
\end{align}
under the $U(3)_{q_L} \times U(3)_u \times U(3)_d$ flavour symmetry of the SM. This is a more restrictive flavour structure than our setup. 

%Instead of coupling the dark sector to right-handed down quark singlets, a coupling Yukawa to up-type quarks or to left-handed quark doublets is allowed by choosing $Y_X = -2/3$ and $Y_X = -1/6$, respectively. In the context of flavoured dark matter, such couplings were recently explored~\cite{Jubb:2017rhm,Blanke:2017fum}. 

\section{Dark meson spectroscopy}
\label{sec:darkmeson}

We assume a hierarchy $m_Q < \Lambda_D$, such that the dark pions, which are the pseudo Nambu-Goldstone bosons of the spontaneously broken dark chiral symmetry, 
are parametrically lighter than other dark hadrons. Heavier composite states such as dark vector-mesons and dark glueballs will undergo fast decays to dark pions.\footnote{An exception are the lightest baryonic bound states, which carry a conserved "dark baryon number" and are therefore stable.} Therefore the phenomenology is largely determined by the lifetimes and decay channels of dark pions. 

The dark pions arise from the $SU(n_d)_L \times SU(n_d)_R \to SU(n_d)_V$ chiral symmetry breaking in the dark sector. For three dark flavours, a theory of the eight resulting dark pions can be written down in analogy to the pions and kaons of QCD. Using the usual basis of Gell-Mann matrices $\lambda^a$,
\begin{equation}
\Pi_D= \pi_D^a \lambda^a = \frac{1}{2} \begin{pmatrix}
\pi_{D3}+\frac{\pi_{D8}}{\sqrt{3}} & ~~~~\pi_{D1}-i\pi_{D2} & ~~~~\pi_{D4}-i \pi_{D5} \\
\pi_{D1}+i \pi_{D2} &~~~~ -\pi_{D3}+\frac{\pi_{D8}}{\sqrt{3}} &~~~~ \pi_{D6}-i \pi_{D7} \\
\pi_{D4}+i\pi_{D5} & ~~~~\pi_{D6}+i \pi_{D7} &~~~~ -\frac{2\pi_{D8}}{\sqrt{3}} \\
\end{pmatrix}\,,
\end{equation}
where the dark quark content of the pions is given in Table~\ref{tab:quarkcontent}. Since we assume identical masses for the dark quarks, the eight dark pions are degenerate in mass, up to tiny splittings induced by their coupling to the SM.

\begin{table}
\begin{center}
\begin{tabular}{c c}
\hline
Dark Pions & Dark quark content \\
\hline
$\pi_{D_1}$, $\pi_{D_2}$ & $\bar{Q}_1 Q_2$, $\bar{Q}_2 Q_1$ \\
$\pi_{D_4}$, $\pi_{D_5}$ & $\bar{Q}_1 Q_3$, $\bar{Q}_3 Q_1$ \\
$\pi_{D_6}$, $\pi_{D_7}$ & $\bar{Q}_2 Q_3$, $\bar{Q}_3 Q_2$ \\
$\pi_{D_3}$, $\pi_{D_8}$ & $\bar{Q}_i Q_i$\\
\hline
\end{tabular}
\end{center}
\caption{Dark quark content of the dark pions.}
\label{tab:quarkcontent}
\end{table}

%\subsection{Decays of dark pions} % I guess we don't really need the subsection
%\label{sec:decays} 
The dark pions can decay into SM hadrons via the Yukawa coupling $\kappa$. Integrating out the heavy mediator field $X$ and performing a Fierz transformation, the decays arise from the dimension-6 effective operator 
\begin{align}
	{\cal L}_{\rm decay} = \frac{\kappa^*_{\alpha i} \kappa_{\beta j}}{2M_X^2} \left( \bar{Q}_\alpha \gamma_\mu P_L Q_\beta \right)\left( \bar{d}_i \gamma^\mu P_Rd_j \right),
\end{align}
upon matching the dark quark current onto a chiral Lagrangian for the dark pions. 
If the $\kappa$ matrix is real, which we will assume in the following, this decay can only happen if the dark pion mass is greater than $3m_{\pi}$, since decays to two SM pions are forbidden by $CP$. Below that threshold, only radiatively induced decays of dark pions into photon pairs or leptons are possible. 
For light dark pions ($m_{\pi_D}\lesssim 4\pi f_{\pi}$), decays are best described using chiral perturbation theory for the SM pions and kaons. For more energetic final states --- that is, larger dark pion masses --- the inclusive decay rate into hadrons can be calculated under the assumption of quark-hadron duality~\cite{Poggio:1975af,Shifman:2000jv}. We follow the simple recipe that if the mass of the dark pions is less than 1.5 GeV, the SM final states are treated using chiral perturbation theory, whereas for larger masses the partonic picture is used.\footnote{The matching between these two pictures at 1.5 GeV works well to within $\mathcal{O}(1)$ factors. See Appendix~\ref{appendix:chpt} for details.}

\begin{figure}
\begin{center}
\begin{subfigure}[b]{0.5\textwidth}
\includegraphics[width=\textwidth]{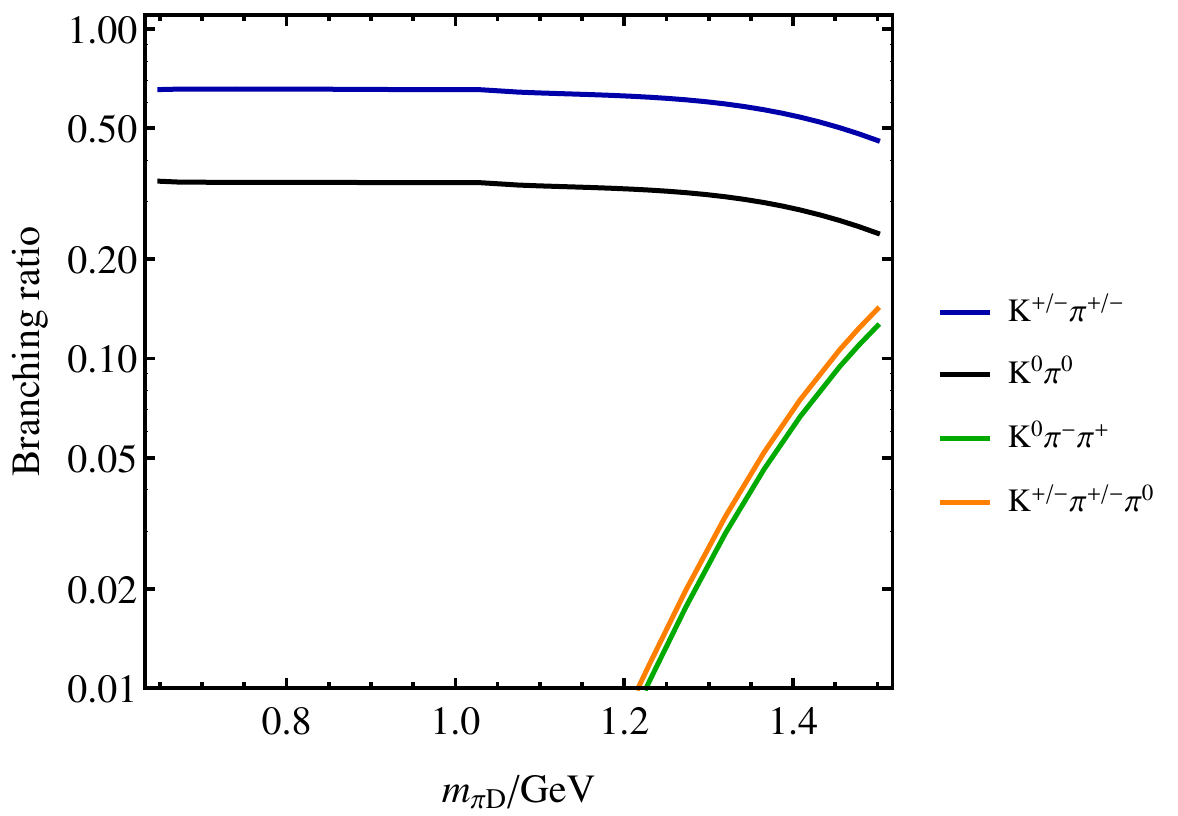}
\caption{$\pi_{D_1}$ and $\pi_{D_2}$}
\end{subfigure}~~~\begin{subfigure}[b]{0.5\textwidth}
\includegraphics[width=\textwidth]{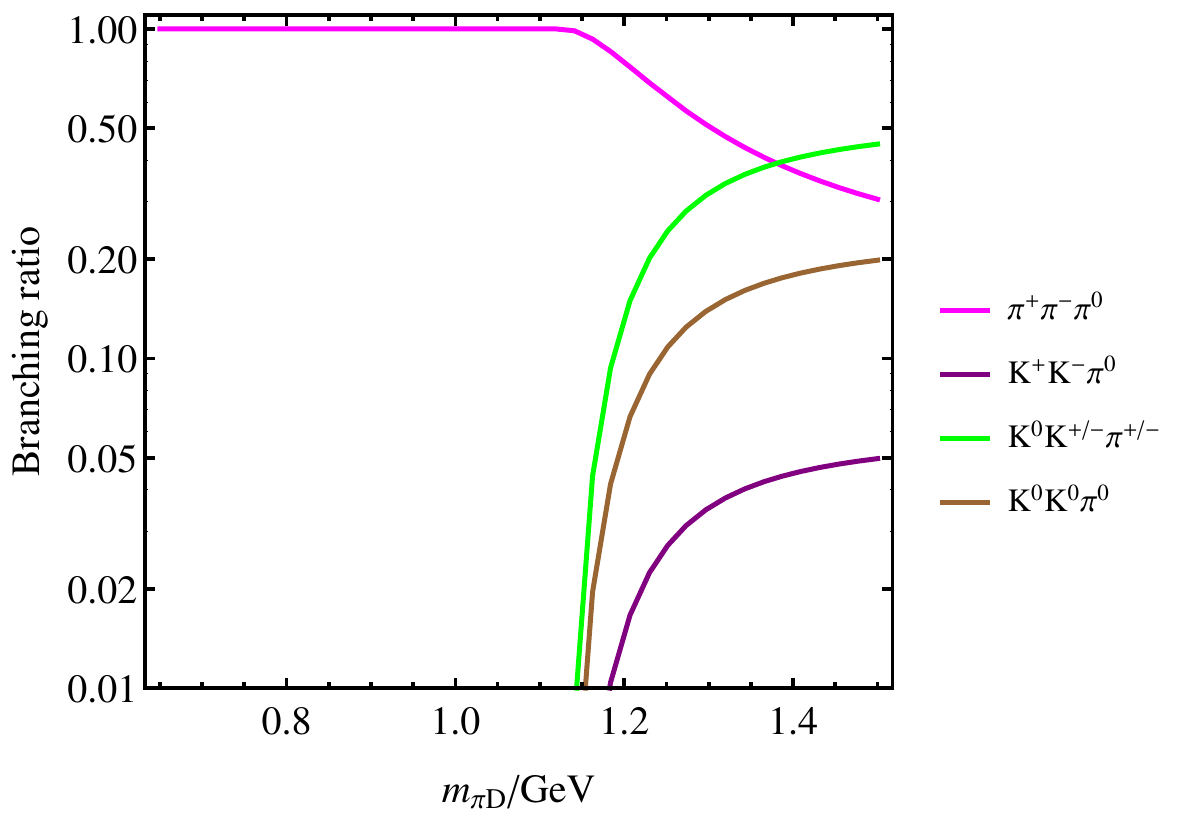}
\caption{$\pi_{D_3}$ and $\pi_{D_8}$}
\end{subfigure}
\caption{Branching ratios of the dark pions $\pi_{D_1}$ and $\pi_{D_2}$, and $\pi_{D_3}$ and $\pi_{D_8}$, for an ``aligned'' coupling matrix $\kappa = \kappa_0 \mathbb{1}_{3\times 3}$. The remaining dark pions, $\pi_{D_4}$, $\pi_{D_5}$, $\pi_{D_6}$ and $\pi_{D_7}$ decay via a loop and/or additional SM flavour breaking, and will be longer-lived.}
\label{fig:lowmasspion1BRs}
\end{center}
\end{figure}

The chiral Lagrangian below 1.5 GeV is 
\begin{align}
\label{eqn:chpt}
\mathcal{L}_{\chi PT}=\frac{f_D^2 f_{\pi}^2}{2 M_X^2} \kappa_{\alpha i}^{*} \kappa_{\beta j} ~\text{tr}(c_{\alpha \beta}U_D^{\dagger} \partial_{\mu} U_D)~ \text{tr}(c_{ij} U\partial^{\mu} U^{\dagger} )\,,
\end{align}
where $f_{\pi}$ is the pion decay constant, $f_D$ is the dark pion decay constant and
\begin{align}
\label{eqn:chiraldefs}
U&=\exp \left[ \frac{2i}{f_{\pi}} \Pi \right],\nonumber \\
U_D &=\exp \left[ \frac{2i}{f_{D}} \Pi_D \right],\\
\Pi&=  \pi^a \lambda^a. \nonumber
\end{align}
The matrices $c_{\alpha\beta}$ are defined as $c_{\alpha \beta}^{mn} \equiv \delta^{m}_{\alpha} \delta^{n}_{\beta}$ ($\alpha$,$\beta$=$1,2,3$), and $c_{ij}^{mn}\equiv \delta^{m}_{i+1} \delta^{n}_{j+1}$ ($i,j$=$1,2$).
This Lagrangian respects the same chiral and dark-chiral symmetries as the partonic Lagrangian, as it must. Calculations of decay rates to SM pions and kaons are outlined in Appendix~\ref{appendix:chpt}.
Branching ratios of the dark pions $\pi_{D_1}$ and $\pi_{D_2}$, and $\pi_{D_3}$ and $\pi_{D_8}$, in the chiral picture below 1.5 GeV, and with an ``aligned'' coupling matrix $\kappa = \kappa_0 \mathbb{1}_{3\times 3}$, are shown in Figure~\ref{fig:lowmasspion1BRs}. It is worth noting that, since a flavour-diagonal dark pion $\bar{Q}_{\alpha} Q_{\alpha}$ can mix through dark QCD interactions into another flavour-diagonal one $\bar{Q}_{\beta} Q_{\beta}$, the lifetimes of the flavour-diagonal dark pions are equal and given by the minimum flavour-diagonal lifetime.

From Figure~\ref{fig:lowmasspion1BRs} it is clear that there are two distinct scenarios for the dark pion decays: for a given dark pion, either all possible final states contain one kaon, or they all contain an even number of kaons. This is due to the unbroken flavour subgroup in this ``aligned'' scenario which ensures that a version of strangeness, under which a dark pion containing $Q_2$ is taken to have strangeness $-1$, remains a good quantum number in dark pion decays. Departures from the alignment limit will generally break this dark strangeness, and allow any given dark pion to decay to any of the final states represented in these plots.

Above 1.5 GeV, the SM current is instead written in terms of quark fields:
\begin{align}
\label{eqn:partoniclagrangian}
\mathcal{L}_{parton} &= \frac{f_D^2}{2 M_X^2} \kappa_{\alpha i}^{*} \kappa_{\beta j} ~\text{tr}(c_{\alpha \beta}U_D^{\dagger} \partial_{\mu} U_D)~ (\bar{d}_{Ri}\gamma^{\mu} d_{Rj}),
\end{align}
giving a decay width for a dark pion composed of $\bar{Q}_{\alpha}Q_{\beta}$ decaying into a pair of quarks $\bar{q}_iq_j$

\begin{equation}
\label{eqn:partonicwidth}
\Gamma_{\alpha \beta ij}=  \frac{N_c m_{\pi_D}f_D^2}{8\pi M_X^4} |\kappa_{\alpha i}\kappa_{\beta j}^*|^2 \left(m_{d_i}^2+m_{d_j}^2 \right)\,\sqrt{\left(1-\frac{(m_{d_i}+m_{d_j})^2}{m_{\pi_D}^2}\right)\left(1-\frac{(m_{d_i}-m_{d_j})^2}{m_{\pi_D}^2}\right)}\, ,
\end{equation}
where $N_c=3$ is a colour factor. Note that the width is proportional to the final state quark masses due to the required helicity flip. Therefore the dark pions will generically decay to  $q\bar{q}$ pairs containing the heaviest accessible quark, unless there is a strongly aligned or hierarchical $\kappa$ matrix which counteracts the mass enhancement.

If a dark pion is prevented altogether from decaying to hadrons, due to kinematic and/or flavour considerations, it will decay to leptons and photons through a loop of SM quarks. The rates of these decays are suppressed compared to typical widths to quarks. For example, 
decays to photons occur via the dimension-5 operator $\pi_D \tilde{F}_{\mu\nu} F^{\mu \nu}$.
The estimated width to photons is
\begin{equation}
\Gamma_{\alpha \beta \to \gamma\gamma} \approx \sum_{k=1}^3 \frac{\alpha_{em}^2}{2304 \pi^3}\frac{f_D^2m_{\pi_D}^3}{M_X^4}\left|\kappa_{\alpha k} \kappa^*_{\beta k} \right|^2.
\end{equation}
In the alignment limit, $\sum_k \kappa_{\alpha k} \kappa^*_{\beta k} \propto \delta_{\alpha \beta}$, such that dark pions which carry non-trivial dark flavour quantum numbers, i.e. for which $\alpha \neq \beta$, are prevented from decaying through this channel. This is not sufficient to guarantee their stability, but their lifetimes will be exceedingly large, since decays have to involve contributions from SM flavour breaking terms, and thus are further suppressed by small CKM matrix elements. 

Phenomenologically, this limit is similar to the case of a $Z'$ mediator. There, some of the dark flavour symmetries remain unbroken, such that flavour off-diagonal dark pions don't decay, resulting in significant amounts of missing energy aligned with the dark jets. Typical jets plus missing energy searches would miss these semi-visible jet signatures due to cuts designed to reduce backgrounds from jet energy mis-measurements, such that a dedicated analysis is required~\cite{Cohen:2015toa,Cohen:2017pzm}.

\section{Constraints from flavour}
\label{sec:flavour}
\tikzset{
    photon/.style={decorate, decoration={snake}},
    fermion/.style={postaction={decorate},
        decoration={markings,mark=at position .55 with {\arrow{>}}}},
    antifermion/.style={postaction={decorate},
        decoration={markings,mark=at position .55 with {\arrow{<}}}},
    gluon/.style={decorate,
        decoration={coil,amplitude=4pt, segment length=4pt}} 
}
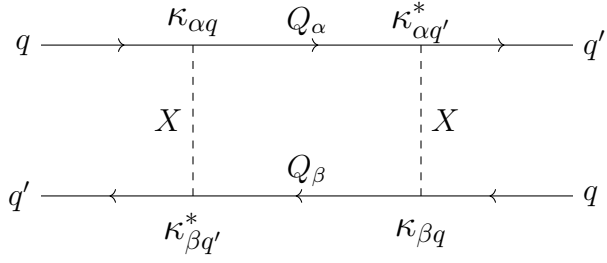
\begin{figure}
\begin{center}
\begin{tikzpicture}
\draw[fermion] (0,2.5) node[left]{\large $q$} -- (2,2.5);
\draw[fermion] (2,2.5) -- (5,2.5)node[midway,above]{\large $Q_{\alpha}$};
\draw[fermion] (5,2.5) -- (7,2.5) node[right]{\large $q^{\prime}$};
\draw[dashed] (2,2.5) -- (2,0.5) node[midway,left]{\large $X$};
\draw[dashed] (5,2.5) -- (5,0.5) node[midway,right]{\large $X$};
\draw[fermion] (2,0.5) -- (0,0.5) node[left]{\large $q^{\prime}$};
\draw[fermion] (5,0.5) -- (2,0.5)node[midway,above]{\large $Q_{\beta}$};
\draw[fermion] (7,0.5) node[right]{\large $q$} -- (5,0.5) ;
\node(a) at (2,2.8) {\Large $\kappa_{\alpha q}$};
\node(b) at (5,2.8) {\Large $\kappa^*_{\alpha q^{\prime}}$};
\node(c) at (2,0) {\Large $\kappa^*_{\beta q^{\prime}}$};
\node(d) at (5,0) {\Large $\kappa_{\beta q}$};
\end{tikzpicture}
\end{center}
\caption{Parton level diagram mediating meson mixing.}
\label{fig:mesonmixing}
\end{figure}
\begin{figure}
\begin{center}
\begin{tikzpicture}
\draw[fermion] (0,2.5) node[left]{\large $q$} -- (2,2.5);
\draw[fermion] (2,2.5) -- (4,2.5)node[right]{\large $Q_{\alpha}$};
\draw[dashed] (2,2.5) -- (2,0.5) node[midway,left]{\large $X$};
\draw[fermion] (2,0.5) -- (0,0.5) node[left]{\large $q^{\prime}$};
\draw[fermion] (4,0.5)node[right]{\large $Q_{\beta}$} -- (2,0.5);
\node(a) at (2,2.8) {\Large $\kappa_{\alpha q}$};
\node(c) at (2,0) {\Large $\kappa^*_{\beta q^{\prime}}$};
\end{tikzpicture}
\end{center}
\caption{Parton level diagram for $B \to K^{(*)}$+ invisible and $K \to \pi$+ invisible.}
\label{fig:invisibledecays}
\end{figure}
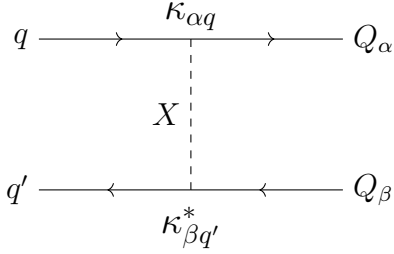
Two types of flavour observables best constrain $\kappa$:
\begin{itemize}
	\item $\Delta F = 2$ processes, in particular $K - \bar{K}$ and $B_{(s,d)} - \bar{B}_{(s,d)}$ mixing, which receive contributions from the box diagrams shown in Figure~\ref{fig:mesonmixing}. 
	\item $\Delta F = 1$ exotic decays of $B$ and $K$ mesons into dark pions, through the diagram shown in Figure~\ref{fig:invisibledecays}.
\end{itemize}
Other flavour measurements such as $b \to s \gamma$ or $b \to s \mu^+\mu^-$ produce constraints subdominant to these~\cite{Agrawal:2014aoa}.

\subsection{Meson mixing}
\label{sec:mesonmixing}
Exchange of $X$ scalars and dark quarks can mediate neutral meson mixing, as seen in Figure~\ref{fig:mesonmixing}. The contributions are proportional to
\begin{align}
	\left(\sum_{\alpha=1}^3  \kappa_{\alpha q}\kappa^*_{\alpha q'} \right)^2\,,
\end{align}
with $q=s$ and $q' = d$ for kaon mixing and $q=b$, $q'=(s,d)$ for neutral $B_{(s,d)}$ meson mixing. It is easy to see that this contribution vanishes in the flavour universal limit, $\kappa_1 = \kappa_2 = 0$, as in this case $\kappa$ is proportional to a unitary matrix: 
\begin{align}
	\left(\sum_{\alpha=1}^3  \kappa_{\alpha q}\kappa^*_{\alpha q'} \right)^2 = \left( [UD (UD)^\dagger]_{qq'} \right)^2 = \kappa_0^4\left( [U U^\dagger]_{qq'}\right)^2=  0 \qquad \text{for $q \neq q'$,}
\end{align}
leaving $\kappa_0$ unconstrained.\footnote{The coupling to the first generation quarks is also constrained by measurements of angular correlations in dijet events at LHC~\cite{Sirunyan:2017ygf,Aaboud:2017yvp,Alte:2017pme}, under the assumption that dark jets are reconstructed as ordinary jets by the LHC experiments. Even then, for TeV scale $M_X$ order one couplings are still allowed. } 
Away from the universal limit, one can see that e.g. if $\kappa_1 = \kappa_2$, the dependence of the mixing amplitude on $U_{12}$ drops out (see Appendix for full calculation), and similarly for cases where the 13 or 23 components of $D$ are degenerate. Thus constraints from $\Delta F = 2$ measurements can be evaded if either all $\theta_{ij}$ are small or if only those $\theta_{ij}$ are large for which the corresponding entries in $D$ are almost degenerate. 

\begin{figure}
\begin{center}
\includegraphics[width=10cm]{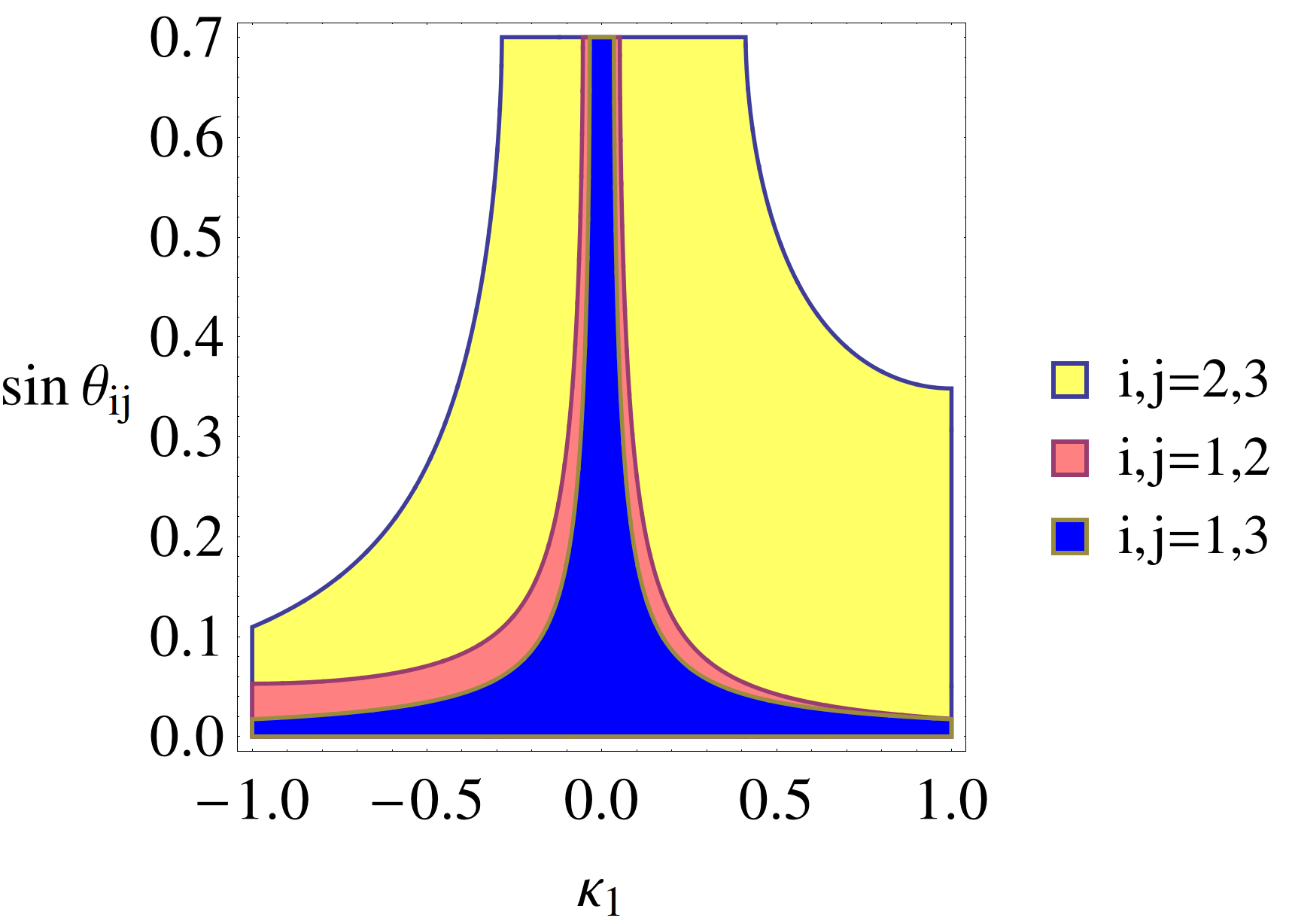} 
\caption{Regions of the $\kappa$ coupling parameter space allowed at 95\% confidence level by measurements of $K^0- \bar{K}^0$ and $B_{(d,s)} - \bar{B}_{(d,s)}$ mixing. The three regions correspond to the sets of parameter choices given in Eqns.~\eqref{params12}-\eqref{params23}.}
\label{fig:mesonmixingbounds}
\end{center}
\end{figure}

We can therefore put constraints on the angles $\theta_{ij}$ and departures from degeneracy in the $i,j$ entries of $D$, with $\Delta_{ij} \equiv D_{ii}-D_{jj}$. Specifically, to derive the constraints shown in Figure~\ref{fig:mesonmixingbounds}, we make the following parameter choices: 
\begin{align}
\label{params12}
\bf{ij=12}:~~~~&\kappa_0=1, && \kappa_1=\kappa_1, && \kappa_2=0,&& \theta_{12}=\theta_{12}, &&\theta_{13}=0, && \theta_{23}=0, \\
\label{params13}
\bf{ij=13}:~~~~&\kappa_0=1, && \kappa_1=\kappa_1, && \kappa_2=0,&& \theta_{12}=0, &&\theta_{13}=\theta_{13}, && \theta_{23}=0, \\
\label{params23}
\bf{ij=23}:~~~~&\kappa_0=1, && \kappa_1=\kappa_1, && \kappa_2=0,&& \theta_{12}=0, &&\theta_{13}=0, && \theta_{23}=\theta_{23},
\end{align}
with $m_X=1$ TeV, $m_Q=2$ GeV,\footnote{But N.B. limits are almost insensitive to the dark quark mass $m_Q$ due to the presence of the much heavier $X$ within the loop.} and the complex phases $\delta_{ij}$ set to zero in every case. The parameter $\kappa_1$ is related to the $\Delta_{ij}$s as $\kappa_1=\Delta_{12}=\Delta_{23}=\Delta_{13}/2$. In calculating the constraints we use the results of Ref.~\cite{Bona:2016bvr} for the new physics (NP) parameter ranges, and adapt the calculations in Ref.~\cite{Agrawal:2014aoa} for our model.

Compared with the analysis of~\cite{Agrawal:2014aoa}, an additional complication in evaluating the numerical constraints coming from neutral meson mixing is that dark gluons can be exchanged between the $Q$ and $X$ fields in Figure~\ref{fig:mesonmixing}. Since $\Lambda_D$ is often above the QCD scale, this introduces a large non-perturbative uncertainty. We try to accommodate this by including a $\pm 50\%$ uncertainty on the NP amplitude, which is included in the regions of Figure~\ref{fig:mesonmixingbounds}. 

\subsection{Exotic decays $K \to \pi \pi_D$ and $B \to K^{(*)} \pi_D$}
\label{sec:BandKdecays}

If the dark pions are light enough to be produced in the decays of $B$ and/or $K$ mesons, and are stable on detector scales, they will contribute to rare decays of these mesons involving missing energy. The processes $K \to \pi \bar{\nu} \nu$ and $B \to K^{(*)} \bar{\nu} \nu$ are suppressed in the SM (see Table~\ref{tab:semiinvisible}), therefore strong constraints on  $\kappa$ arise from these measurements. 

\begin{table}
\begin{center}
\begin{tabular}{c c c}
\toprule
Observable & Measurement or bound & SM prediction \\
\midrule
${\rm Br} (K^+ \to \pi^+ \bar{\nu} \nu)$  & $\left(17.3^{+11.5}_{-10.5}\right)\times 10^{-11} ~\text{\cite{Artamonov:2008qb}}$ & $\left(8.4\pm 1.0\right)\times  10^{-11}~\text{\cite{Buras:2015qea}}$ \\
$ {\rm Br} (B^+ \to K^+ \bar{\nu} \nu)$ & $< 1.7 \times 10^{-5}~ \text{\cite{Lees:2013kla}}$ & $\left(4.0\pm 0.5\right)\times  10^{-6} ~\text{\cite{Buras:2014fpa}}$\\
${\rm Br} (B^0 \to K^{*0} \bar{\nu} \nu)$ &$< 5.5 \times 10^{-5} ~\text{\cite{Lees:2013kla}}$ &  $\left(9.2\pm 1.0\right)\times  10^{-6}~\text{\cite{Buras:2014fpa}}$\\
\bottomrule
\end{tabular}
\end{center}
\caption{Measurements and experimental bounds on the branching ratios of semi-invisible meson decays, and their SM predictions. Bounds are reported at $90\%$ confidence level.}
\label{tab:semiinvisible}
\end{table}

%

%then strong constraints can be put on the $\kappa$ coupling from $K \to \pi \bar{\nu} \nu$ and $B \to (K^{(*)}, \pi) \bar{\nu} \nu$ measurements. 

The decays are induced by the dimension-5 operator
\begin{equation}
%\mathcal{O}_{(d_i\to d_j \pi_D)}=
\sum_{\alpha, \beta} \kappa_{\alpha i} \kappa^*_{\beta j} \frac{f_{\pi_D}}{M_X^2} \left(\bar{d}_R^i \gamma^{\mu} d_R^j \right)\partial_{\mu} \pi_D,
\end{equation}
for decays involving a single dark pion (eg.~$B \to K \pi_D$), or the dimension-6 operator 
\begin{equation}
%\mathcal{O}_{(d_i\to d_j \bar{Q} Q)}=
\sum_{\alpha, \beta} \kappa_{\alpha i} \kappa^*_{\beta j} \frac{1}{2 M_X^2} \left(\bar{d}_R^i \gamma^{\mu} d_R^j\right)\left(\bar{Q}^{\beta}\gamma_{\mu} Q^{\alpha} \right),
\end{equation}
for decays involving an open dark quark pair (e.g.~$B \to K \bar{Q} Q$).
Here $i$, $j$ are quark flavours and $\alpha$, $\beta$ are dark quark flavours. In the limit that $\kappa_1$ and $\kappa_2$ are small compared to $\kappa_0$, the strength of this interaction depends only on $\kappa_0$, and not on the $\theta$ and $\delta$ parameters which drop out due to unitarity. 

\begin{figure}
\begin{center}
\includegraphics[width=11cm]{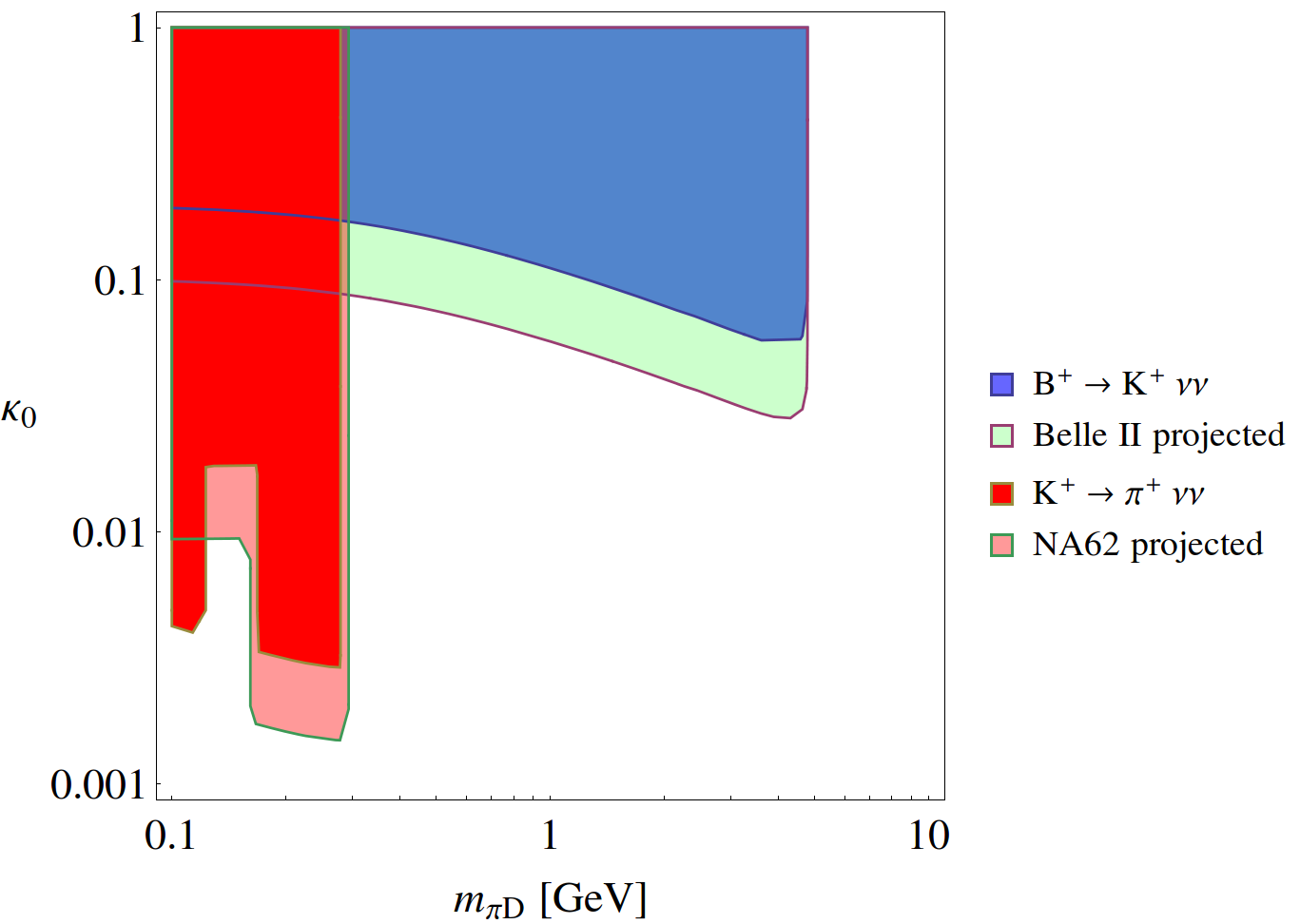} 
\caption{Regions excluded by current limits of semi-invisible meson decays (see Table~\ref{tab:semiinvisible}) and projected limits from the upcoming Belle II and NA62 experiments, in the ``aligned'' flavour scenario in which $\kappa=\kappa_0 \mathbb{1}_{3\times 3}$.}
% In black are points that are not excluded by the constraints from meson mixing and semi-invisible decays, when the parameters of the $\kappa$ matrix are sampled from the following ranges: $\kappa_1, \kappa_2= [-\kappa_0, \kappa_0]$, $\theta_{ij}=[0,\pi/4]$. }
\label{fig:scatterflavour}
\end{center}
\end{figure}

In this case, (or alternatively if $\kappa_1$, $\kappa_2$, $\theta_i$ are given fixed values), experimental limits can be phrased as bounds on $\kappa_0$ and the various mass scales $M_X$, $m_{\pi_D}$, $f_D$ and $m_Q$. 
Taking the relevant branching ratio expressions from Ref.~\cite{Kamenik:2011vy}, the current bounds from $B^+ \to K^+ \bar{\nu} \nu$~\cite{Lees:2013kla} (blue) and $K^+ \to \pi^+ \bar{\nu} \nu$~\cite{Artamonov:2008qb} (red) are shown in Fig~\ref{fig:scatterflavour}, taking $f_{\pi_D}=m_{\pi_D}=10m_Q$, and $\kappa=\kappa_0 \mathbb{1}$. Both the dark pion channel and the open dark quark channel are included in the calculation of the bounds, although in practice the open dark quark channel only becomes important at very low $m_{\pi_D}$ (or in regions where the single dark pion channel is removed by experimental cuts). The strange feature in the $K^+ \to \pi^+ \bar{\nu} \nu$ limit is due to the fact that, in order to reduce backgrounds, the measurement of $K^+ \to \pi^+ \bar{\nu} \nu$ was done within two regions of the invariant mass $q^2$ of the invisible particles, $q^2/m_K^2 \in [0.00,0.062]$  and $q^2/m_K^2 \in [0.116,0.289]$. Following the approach taken in Ref.~\cite{Kamenik:2011vy}, we find conservative bounds by demanding that the NP contribute no more than $10^{-10}$ branching ratio when summed over both experimental regions. The effects of the experimental cuts are that the excluded region only extends to $2m_{\pi}$, and if the mass of the dark pion falls between the two regions, then only the open dark quark channel can contribute. 

Although the excluded regions in Figure~\ref{fig:scatterflavour} have been calculated with particular assumptions on $\kappa$, the bounds are rather insensitive to changing the parameters. However, if $\kappa_1$ and/or $\kappa_2$ are chosen such that one of the entries of the $D$ matrix completely disappears (or becomes very small), then one or both of these bounds can be evaded, since in this case the coupling of the dark sector to one of the quarks vanishes.

These bounds of course should only apply if (enough of) the dark pions are stable on detector scales, hence mimicking neutrinos in the relevant experiments. The question is whether there is an allowed region for large $\kappa_0^2 f_{\pi_D}/M_X^2$ in which the dark pions decay quickly enough to be unconstrained by these limits. But it turns out that if $\kappa_0^2 f_{\pi_D}/M_X^2$ becomes large enough that the proper decay length $c\tau_0$ of any of the dark pions is of order of metres or below ($\kappa_0 \gtrsim 0.25$, depending on the mass), the decays producing dark pions are contributing around 10\% or more of the total $B$-meson decay width. Since the bounds on the branching ratios ${\rm Br}( B \to K^{(*)} \bar{\nu} \nu)$ are $O(10^{-5})$, and given that some fraction of the dark pions will escape the detector, the bounds still apply. (Not to mention that for these huge widths it is likely that the dark pions that do decay within the detector would produce noticeable effects in other $B$ branching ratio measurements.) In fact, for the particular case of the aligned scenario $\kappa = \kappa_0 \mathbb{1}$, the dark pions $\pi_{D_6}$ and $\pi_{D_7}$ require additional loops and SM flavour breaking to decay, and will always have very long decay lengths.

While these meson decay constraints severely limit the magnitude of $\kappa$ in the case of low confinement scales in the dark sector, there are some welcome consequences. First, in coming years, the NA62 experiment will measure ${\rm Br}(K^+ \to \pi^+ \bar{\nu} \nu)$ to within 10\% of the SM value~\cite{Martellotti:2015kna}, while Belle II should be sensitive to the SM $B \to K^{(*)} \bar{\nu} \nu$ branching ratios at 30\% accuracy with 50ab$^{-1}$ of data~\cite{Buras:2014fpa,Aushev:2010bq,Altmannshofer:2009ma}. These will provide opportunities to either discover or further constrain the model. The projected reach of these measurements is shown in Figure~\ref{fig:invisibledecays}.
Furthermore heavy flavour mesons are produced ubiquitously at fixed target experiments, and therefore these decays can contribute to the total dark pion yield. In fact they will turn out to be the dominant source of dark pions in the region of parameter space where those decays are allowed, as we discuss in more detail in Sec.~\ref{sec:fixedtarget}.

\section{Cosmology}
\label{sec:cosmology}
The dark matter in our model consists of dark baryons -- bound states of dark quarks with unit dark baryon number. Since the annihilation cross section $p_D \bar{p}_D \to \pi_D, \rho_D, \dots$ is much larger than the one required by the freeze-out mechanism, a non-vanishing DM relic abundance only survives if an asymmetry between dark baryons and dark anti-baryons is induced in the early universe. This can be achieved in several ways for the particle content considered here, e.g.~\cite{Blennow:2010qp,Frandsen:2011kt,Bai:2013xga}, and we assume that one such mechanism is implemented at some higher scale. Both baryon and dark baryon number are conserved separately by our model at the TeV scale, so no additional constraints arise from requiring that the DM abundance is not washed out. 

Thermal equilibrium between the visible and dark sectors is established at high temperatures $T > M_X$ by QCD and dark QCD interactions alone, independent of the values of the Yukawa couplings.\footnote{This is the case because processes like $gg \to X\bar{X}$ thermalise $X$, which in turn allows scatterings of the form $g X \to X g_D$ to thermalise the dark gluons, which then equilibrate the rest of the dark sector.}
Below $M_X$, the process $gg \to g_D g_D$ is described by a loop induced dimension 8 operator $G^2 G_D^2/\Lambda^4$ with $\Lambda^{-4} \sim \alpha_s \alpha_D/M_X^4$. It is therefore strongly suppressed at lower temperatures, and fails to maintain equilibrium below $T \approx 15$~GeV. 
For $\kappa \gtrsim 0.03$, Yukawa mediated scatterings $q \bar{q} \to Q \bar{Q}$ can keep both sectors in equilibrium down to the GeV scale, where the parton level picture becomes invalid. 

Once the temperature drops below $\Lambda_D$, the dark sector will consist of a thermal bath of mostly dark pions with temperature close to that of the visible sector. Decays of these dark pions to SM particles will eventually transfer back the entropy to the visible sector, leaving only the stable dark matter behind. An important constraint is that this entropy transfer should not disrupt big bang nucleosynthesis (BBN).

\subsection{BBN constraints}
Nucleosynthesis,~i.e. the formation of light elements out of a thermal bath of protons and neutrons at $T < {\rm MeV}$, is very sensitive to the injection of energy from late decaying particles. Once BBN has ended, very light decays still affect the ratio of photon to neutrino temperature and would therefore most likely be in conflict with the number of relativistic degrees of freedom at the time when the CMB forms, which is determined accurately from Planck and WMAP data~\cite{Ade:2015xua}. To avoid these constraints it is sufficient to require that all unstable particles have lifetimes of less than one second. 

We have seen above that even for ${\cal O}(1)$ values of the Yukawa couplings, some dark pions can be very long lived due to accidental flavour symmetries, and thus potentially in conflict with BBN.
However here the situation is slightly more complicated. Consider two dark pions, $\pi_{Ds}$ with $\Gamma^{-1} < 1$~s, and $\pi_{Dl}$ with $\Gamma^{-1} > 1$~s. Dark chiral perturbation theory then gives
\begin{align}
	\langle \sigma v \rangle ( \pi_{Dl} \pi_{Dl} \to \pi_{Ds} \pi_{Ds}) \sim \frac{T^2}{f_{\pi_D}^4}\,,
\end{align}
for the thermally averaged cross section. If these processes are still in equilibrium once $\pi_{Ds}$ starts to decay, then the $\pi_{Dl}$ abundance is depleted along with $\pi_{Ds}$, and BBN is safe. This is clearly the case down to temperatures below BBN. The condition for evading the BBN constraints is therefore relaxed -- is is only necessary that one of the dark pions has a lifetime shorter than one second, which is in general satisfied for dark pion masses above one GeV and coupling $\kappa_0\gtrsim 0.01$. The region disfavoured by BBN is shaded grey in Figure~\ref{fig:ship}.

\subsection{Dark matter direct detection}
The lightest dark baryon, prevented from decaying by dark baryon number conservation, is a dark matter candidate. Limits can be put on its mass and interactions from direct detection experiments.

Since the masses of the dark quarks are assumed to be degenerate, we have eight degenerate dark baryons $p_{D_k}$ ($k=1,...,8$), analogous to the baryon octet of QCD formed of the $u, d, s$ quarks.
For the dominant spin-independent scattering, the matrix element for scattering of any of these off a proton or a neutron is given by~\cite{Goodman:1984dc} 
\begin{equation}
\mathcal{M}_{p,n}= \sum_{\alpha} \frac{|\kappa_{\alpha 1}|^2}{8 M_X^2} J^{0}_{D\alpha} J^0_{p,n}\,,
\end{equation}
where $J^0_{D\alpha}= \sum_k \left\langle p_{D_k} |\bar{Q}_{\alpha} \gamma^0 Q_{\alpha}| p_{D_k} \right\rangle$ 
and $J^0_{p,n}=\left\langle p,n |\bar{d} \gamma^0 d| p,n \right\rangle \approx 1,2$. When summed over all the dark baryons, $J^0_{D\alpha}=1$, since it corresponds to the number of valence $Q_\alpha$ quarks in all the eight baryons, averaged over the number of baryons.
The averaged spin-independent dark baryon-nucleon cross-section is then~\cite{Bai:2013xga}
\begin{align}
\sigma_{N-D}^{SI}&=\frac{1}{A^2} \sum_{\alpha} \frac{(J^0_{D\alpha})^2|\kappa_{\alpha 1}|^4\mu^2_{n-D}}{32\pi M_X^4} \left( J^0_{n} (A-Z) + J^0_{p} Z \right)^2 \\
&=\frac{1}{A^2} \sum_{\alpha} \frac{|\kappa_{\alpha 1}|^4\mu^2_{n-D}}{32\pi M_X^4}\left(2(A-Z)+Z \right)^2,
\end{align}
where $\mu_{N-D}$ is the reduced mass of the dark baryon-nucleon system. For Xenon, $Z=54$ and $A=131$. The current strongest bounds on dark matter with masses of a few GeV or above comes from the XENON1T experiment~\cite{Aprile:2017iyp}. Direct detection constrains in particular the region of parameter space that can be probed by the emerging jets search~\cite{Schwaller:2015gea}, as can be seen in Figure~\ref{fig:emerging} in the following section.

\section{LHC phenomenology}
\label{sec:emerging}

The heavy coloured $X$ mediators can be pair-produced at the LHC, each one producing a SM jet and a jet of dark hadrons in its decay. These dark hadrons will decay to dark pions, and stable dark protons; the dark pions will then decay back into SM states via the dark portal interaction. Depending on the lifetimes of the dark pions, there are several possible scenarios for how these dark jets appear at the LHC. 

If the lifetimes of the dark pions are short, the dark jet will decay promptly to SM hadrons, and the event will appear as a high energy multi-jet event.\footnote{The stable dark protons within the jets will escape the detector, leading to some missing energy. In the large $N_c$ limit, the production of baryons is suppressed relative to mesons~\cite{Witten:1979kh}, and happens at the 10\% level in QCD~\cite{Beringer:1900zz}. Therefore we expect roughly $10\%$ missing energy in a dark jet, if all the dark pions decay within the detector.}
On the other extreme, if the dark pions are long-lived, such that they escape the detector completely without decaying, the event will appear as a dijet event with missing energy.
In an intermediate scenario, where the decay lengths of (at least some of) the dark pions are of the order of centimetres, the dark jet ``emerges'' into SM states over detectable scales. This is the emerging jets scenario studied in Ref.~\cite{Schwaller:2015gea}.\footnote{See also \cite{Han:2007ae,Strassler:2008fv,Carloni:2010tw,Carloni:2011kk} for earlier studies of dark sector phenomenology at hadron colliders, and~\cite{Knapen:2016hky,Cui:2016rqt,Beauchesne:2017yhh,Park:2017rfb} for recent related work.}

As discussed above, the flavour structure of the dark portal coupling, and constraints on its parameters, strongly affect the rates of the dark pion decays into different SM final states. These considerations determine how the dark jets will appear at the LHC. 

\begin{table}
\begin{tabular}{c  c c c c}
\toprule
\bf{Scenario} & Dark pion & $c\tau_0\kappa_0^4~/mm$ &$c\tau_0\kappa_0^4~/mm$ &$c\tau_0\kappa_0^4~/mm$\\
~ & ~~ & ~($m_{\pi_D}=1$ GeV)~ & ~($m_{\pi_D}=5$ GeV)~ & ~($m_{\pi_D}=10$ GeV) \\
\midrule
\midrule
\bf{``Aligned''} & $\pi_{D_3}$, $\pi_{D_8}$  & $1.4\times 10^4$ & $5.7$ & $0.72$\\
$\sin \theta_{ij}=0$& $\pi_{D_1}$, $\pi_{D_2}$ & $1.4\times 10^2$ & $5.7$ &$0.72$ \\
$\Delta_{ij}=0$& $\pi_{D_4}$, $\pi_{D_5}$ & Long-lived & Long-lived & $4.0\times 10^{-4}$\\
& $\pi_{D_6}$, $\pi_{D_7}$ & Long-lived &  Long-lived & $4.0 \times 10^{-4}$\\
\midrule
\bf{``12''} & $\pi_{D_3}$, $\pi_{D_8}$ & $2.7 \times 10^3$ &$5.6$ & $0.70$ \\
$\sin \theta_{12}=0.1,$& $\pi_{D_1}$, $\pi_{D_2}$ & 62 & $2.6$ &$0.32$\\
$\Delta_{12}=0.5\kappa_0$ & $\pi_{D_4}$, $\pi_{D_5}$ & Long-lived & Long-lived &$7.2 \times 10^{-4}$\\
& $\pi_{D_6}$, $\pi_{D_7}$ & Long-lived & Long-lived &$1.6 \times 10^{-3}$ \\
\midrule
\bf{``13''} & $\pi_{D_3}$, $\pi_{D_8}$ & $5.7 \times 10^3$ & $5.7$ & $3.3 \times 10^{-2}$\\
$\sin \theta_{13}=0.05,$&$\pi_{D_1}$, $\pi_{D_2}$ & 90 & $3.7$ &$8.5 \times 10^{-2}$ \\
$\Delta_{13}=0.5\kappa_0$& $\pi_{D_4}$, $\pi_{D_5}$ & $7.1 \times 10^6$ & $3.7\times 10^5$ &$4.6 \times 10^{-4}$ \\
& $\pi_{D_6}$, $\pi_{D_7}$ & $9.0 \times 10^5$ & $1.5\times 10^{3}$ &$7.2 \times 10^{-4}$ \\
\midrule
\bf{``23''} &  $\pi_{D_3}$, $\pi_{D_8}$ & $2.7 \times 10^3$ & $6.9$ &$2.5 \times 10^{-3}$ \\
$\sin \theta_{23}=0.3,$ & $\pi_{D_1}$, $\pi_{D_2}$  & 68 &$2.8$ & $2.0\times 10^{-3}$\\
$\Delta_{23}=0.5\kappa_0$&$\pi_{D_4}$, $\pi_{D_5}$ & $1.1 \times 10^{4}$ &$1.2 \times 10^2$ & $7.9 \times 10^{-4}$\\
& $\pi_{D_6}$, $\pi_{D_7}$ & Long-lived &$2.9\times 10^{2}$ & $1.9 \times 10^{-3}$\\
\midrule
\bf{``Emerging jets''} & All & $3.0 \times 10^2$ & 17 & $9.1 \times 10^{-4}$ \\
\bottomrule
\end{tabular}
\caption{Proper decay lengths ($c \tau_0$) of the dark pions, for various flavour scenarios (see text for details). ``Long-lived'' means that tree-level decays to hadrons are impossible for these parameter choices, so the dark pion in question will decay via loop induced processes, and is detector stable. The following parameters are used throughout: $m_X=1$ TeV, $f_{\pi_D}=m_{\pi_D}$.}
\label{tab:decaylengths}
\end{table}

Proper decay lengths ($c\tau_0$) for a number of benchmark parameter points are given in Table~\ref{tab:decaylengths}. The numbered $``ij"$ scenarios are as defined in Eqns.~\eqref{params12}-\eqref{params23}, but now leaving $\kappa_0$ free as an extra handle on the decay length. For each of these scenarios, $\Delta_{ij}$ is chosen to be $0.5 \kappa_0$, and $\sin \theta_{ij}$ is taken to be the largest value allowed by meson mixing (when $\kappa_0$=1), without inflating the theoretical errors. The ``emerging jets'' row gives the decay length in a scenario in which all eight dark pions have the same lifetime, to allow comparison with the analysis of Ref.~\cite{Schwaller:2015gea}. In our parameterisation this is achieved by setting
\begin{equation}
\kappa_{EJ} = \frac{\kappa_0}{\sqrt{3}}\begin{pmatrix}
1 & 1 & 1 \\ 1 & 1 & 1 \\1 & 1 & 1 
\end{pmatrix}.
\end{equation}

\begin{figure}
\begin{center}
\includegraphics[width=.6\textwidth]{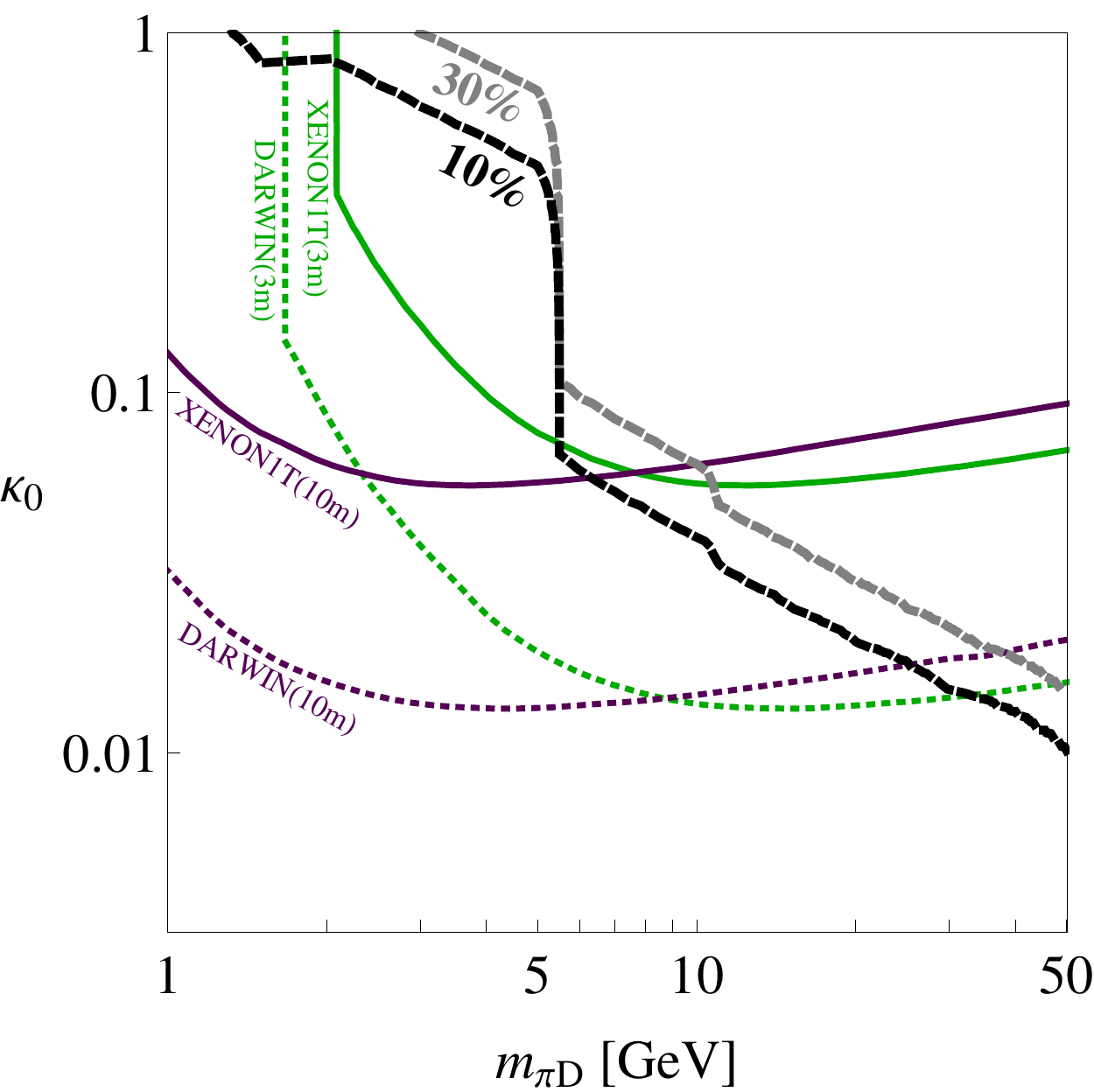} 
\caption{The thick black (grey) line delineates the region in which more than 10\% (30\%) of dark pions have decayed within 1m transverse distance from the beamline at the 14 TeV LHC. Overlaid are current (solid) and projected (dashed) dark matter direct detection bounds, assuming a dark proton mass of  $3 m_{\pi_D}$ (green) or $10 m_{\pi_D}$ (purple). 
This parameter space corresponds to the ``aligned'' scenario, $\kappa=\kappa_0 \mathbb{1}_{3\times 3}$, and choosing $m_X=1$ TeV, $f_{D}=m_{\pi_D}$.}
\label{fig:emerging}
\end{center}
\end{figure}

While Table~\ref{tab:decaylengths} shows that there are flavour safe scenarios for which all dark pions have very similar lifetimes, it is also clear that in general this is not a valid assumption, in particular for $m_{\pi D} \lesssim 5$~GeV. In order to understand the region of parameter space that can be probed with emerging jets searches, we have to require that at least a fraction of the dark pions decays within the inner part of the LHC detectors. 

In Figure~\ref{fig:emerging}, for the aligned scenario $\kappa=\kappa_0 \mathbb{1}_{3\times 3}$, the regions for which the dark pion decay lengths are within detector scales are indicated, along with current and future dark matter direct detection bounds. The solid green and purple lines are current direct detection constraints from XENON1T~\cite{Aprile:2017iyp}, under two different assumptions for the mass of the dark proton $m_{p_D}$: the green line takes $m_{p_D}=3m_{\pi_D}$; the purple line takes $m_{p_D}=10m_{\pi_D}$. The dotted green and purple lines are projected bounds from the proposed DARWIN experiment~\cite{Aalbers:2016jon}, under the same mass assumptions.
Above the grey (black) dashed line is the region for which more than 30\% (10\%) of dark pions produced at the 14 TeV LHC have decayed within 1m transverse distance of the interaction point. In calculating these regions, the pair production of $X$ was simulated with MadGraph~5 \cite{Alwall:2014hca}, using FeynRules \cite{Alloul:2013bka} to implement the model. The decay of the $X$ and the subsequent parton showering and dark parton showering and hadronisation were performed using PYTHIA 8 \cite{Sjostrand:2007gs}.

Independent of the exact dark matter -- dark pion mass relation, we observe that the region of pion masses above $\sim 5$~GeV, which is accessible to LHC searches, will independently be probed by dark matter direct detection experiments in the future. At these masses, decays with final states involving $b$ quarks are accessible kinematically. Due to the quark mass factor appearing in Equation~\eqref{eqn:partonicwidth}, the dark pions will tend to have a large branching ratio to $b \bar{b}$, $b \bar{q}$ and $q \bar{b}$ pairs ($q=d,s$), and hence these ``emerging jets" will contain a large number of $B$-hadrons. Furthermore, if the structure of the $\kappa$ coupling prevents some of the eight dark pions from decaying into $b$-quarks (e.g. $\pi_{D_1}$ and $\pi_{D_2}$ in the ``Aligned" and ``12" scenarios), two distinct decay lengths will be important: a shorter length at which many $b$-flavoured hadrons will emerge, and a longer length at which mostly light- and strange-flavoured hadrons will emerge.

%For smaller dark pion masses, we will see below that fixed target experiments can cover a large portion of the parameter space that is consistent with flavour physics and cosmology. 
%
%It can be seen from Figure~\ref{fig:emerging} that the non-excluded regions for which the dark jets will appear to ``emerge" at the LHC have dark pion masses larger than about 8-10 GeV and coupling strength $\kappa_0$ smaller than about 0.07. At these masses, decays with final states involving $b$ quarks are accessible kinematically. Due to the quark mass factor appearing in Equation~\eqref{eqn:partonicwidth}, the dark pions will tend to have a large branching ratio to $b \bar{b}$, $b \bar{q}$ and $q \bar{b}$ pairs ($q=d,s$), and hence these ``emerging jets" will contain a large number of $B$-hadrons. Furthermore, if the structure of the $\kappa$ coupling prevents some of the eight dark pions from decaying into $b$-quarks (e.g. $\pi_{D_1}$ and $\pi_{D_2}$ in the ``Aligned" and ``12" scenarios), two distinct decay lengths will be important: a shorter length at which many $b$-flavoured hadrons will emerge, and a longer length at which mostly light- and strange-flavoured hadrons will emerge.

\begin{figure}
\begin{center}
\includegraphics[width=.45\textwidth]{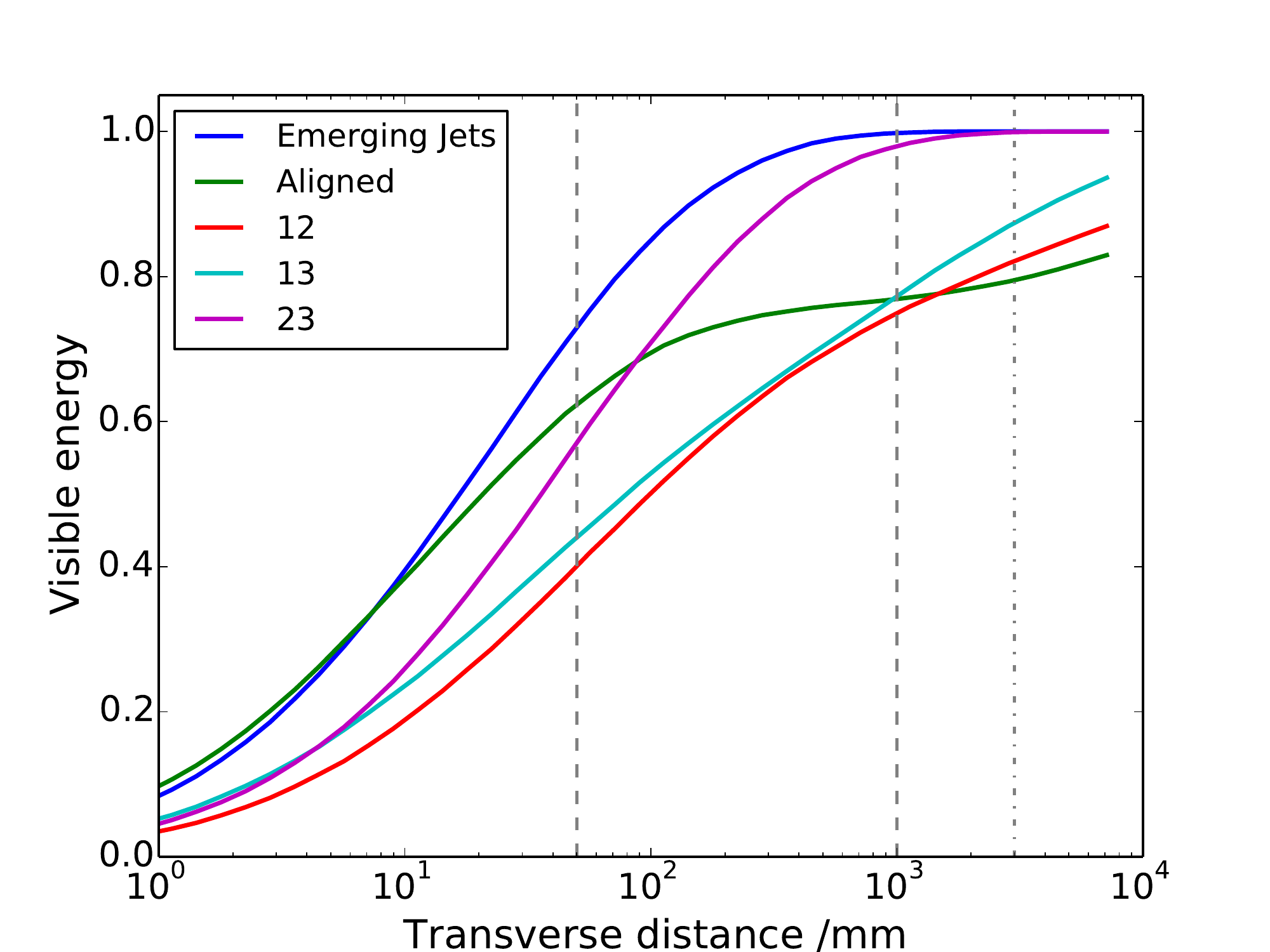} 
\includegraphics[width=.45\textwidth]{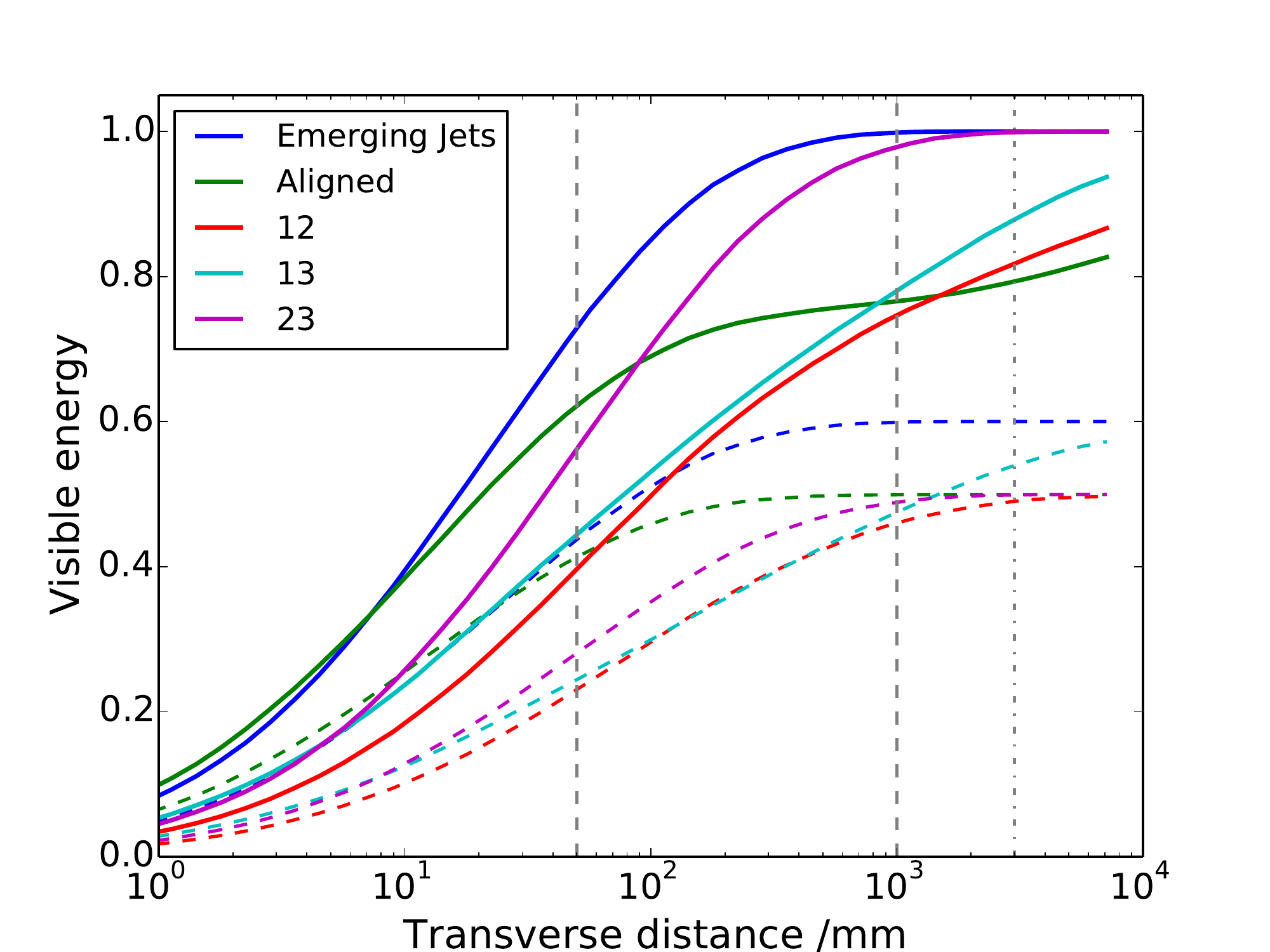} 
\caption{Average visible energy (defined as the energy transferred to SM states when dark pions decay) as a function of transverse distance from the interaction point at the 14 TeV LHC. The flavour scenarios corresponding to the different lines are outlined in Table~\ref{tab:decaylengths} and described in the text. On the right we show, in addition, the fraction of energy in heavy flavours ($b$-jets) as dashed lines. The parameters chosen here are $m_{\pi_D}=20$ GeV, $f_D=m_{\pi_D}$, $\kappa_0$=0.09, $m_X=1$ TeV. }
\label{fig:energydist}
\end{center}
\end{figure}

This behaviour can be seen in Figure~\ref{fig:energydist}. Here we demonstrate the dependence on the flavour scenario of the ``emerging" nature of the jet at the 14 TeV LHC, by plotting the average ``visible energy'' --- meaning the energy transferred to SM states when the dark pions decay --- against the transverse distance from the beamline in millimetres. The energy is normalised to the total energy carried by dark pions in the dark jet. The parameters chosen are $m_{\pi_D}=20$ GeV, $\kappa_0=0.09$, which lead to dark jets with decay lengths of the order of LHC detector scales. It can be seen most clearly for the ``aligned'' scenario that there are two rather different decay lengths, such that although many decays occur between centimetre to metre scales, by around 1m the number of decays has levelled off at a point where only $\sim 75\%$ of the energy carried by the dark pions has been converted into SM particles. The remaining dark pions have a longer decay length and will decay outside the detector -- in fact the visible energy can be seen to begin to grow again at distances of order 10m. As discussed above, this is characteristic of flavour scenarios in which some dark pions are prevented from decaying to $b$ quarks. 

In the plot on the right hand side of Figure~\ref{fig:energydist}, the dotted lines represent the visible energy carried by $b$ quarks. It can be seen from the slopes of these lines (again, most clearly for the ``aligned'' scenario) that the decays involving $b$ quarks are responsible for the shorter decay lengths within the dark jet, since they reach a maximum and level off over centimetre to metre length scales, while other decays continue to occur up to tens of metres and beyond.

\section{Fixed target experiments}
\label{sec:fixedtarget}

At fixed target experiments such as the running NA62 \cite{NA62:2017rwk} and the proposed SHiP \cite{Alekhin:2015byh} experiments at the CERN SPS, the main source of dark pions will be in decays of $B$ mesons\footnote{Direct production of dark quarks via $t$-channel exchange of the scalar $X$ is also possible, but (for $X$ masses around the TeV scale) this production mode is suppressed by several orders of magnitude as compared to production through $B$ decays.} (see Sec.~\ref{sec:BandKdecays}). 

\begin{figure}
\begin{center}
\includegraphics[width=12cm]{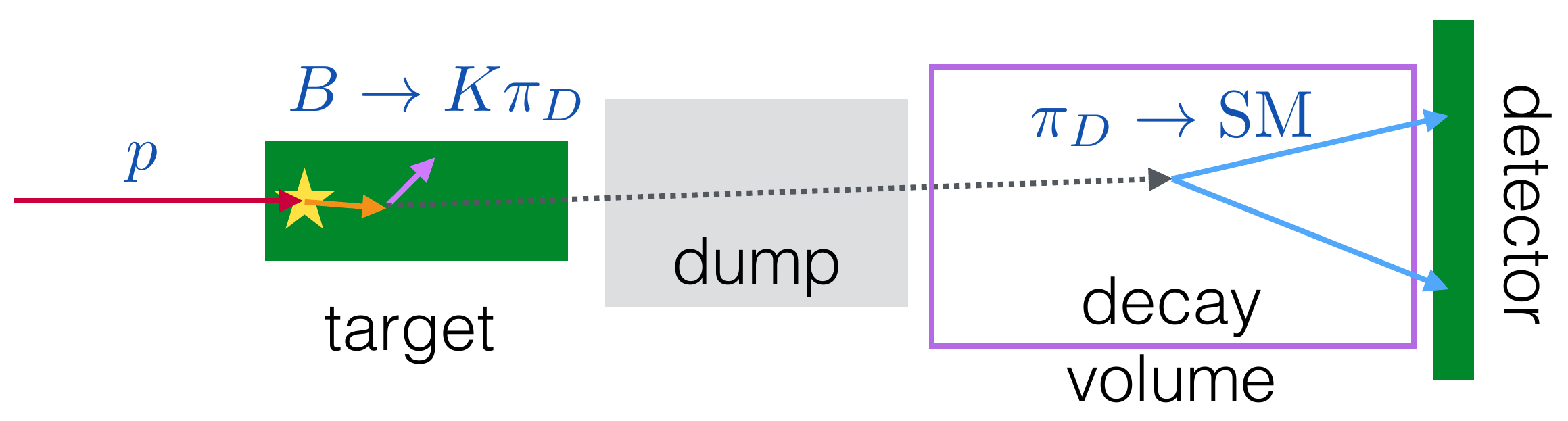} 
\caption{Schematic view of dark pion production and decay in a fixed target experiment. A 400~GeV proton has a probability of $1.6\times 10^{-7}$ to produce a pair of $b$-quarks~\cite{Alekhin:2015byh} when scattering off the target. With $10^{20}$ ($10^{18}$) protons on target, up to $10^{8}$ ($10^{6}$) dark pions can be produced from $B$ meson decays in the SHiP and NA62 experiments, respectively. 
%Detectability further depends on the geometric acceptance, lifetime and decay channels of the dark pions. 
}
\label{fig:fixedt}
\end{center}
\end{figure}

The design of the proposed SHiP experiment comprises a tungsten target, followed by a hadron absorber and a muon shield before a 50m long decay vessel beginning 60m from the target. Hidden sector particles, produced when the 400 GeV proton beam hits the target, may pass through the absorber and shield and decay within the decay volume. The expected number of proton-target collisions is $2 \times 10^{20}$ over 5 years of operation and the total number of $B$ mesons produced is foreseen as $7 \times 10^{13}$. Following the calculations of Sec.~\ref{sec:BandKdecays}, some of these can decay to a $K^{(*)}$ meson (or a pion) and a dark pion. The $K^{(*)}$ will be stopped in the hadron absorber, but the dark pion may pass through to the decay volume, where it may decay to pions and kaons. 

The total number of dark pion decays expected in the SHiP decay volume is estimated as
\begin{equation}
N_{\pi_D} = N_B \cdot {\rm Br}(B \to K^{(*)} \pi_D) \cdot \epsilon_{geom} \cdot F_{decay},
\end{equation}
where $N_B$ is the number of $B$ mesons produced ($N_B= 7 \times 10^{13}$). The geometric acceptance $\epsilon_{geom}$ is defined as the fraction of dark pions with lab-frame momentum at an angle $\theta < \theta_{max}= 2.5/60$ from the beam axis, such that they pass into the 5m-diameter decay volume \cite{Alekhin:2015byh}. The fraction of dark pions that decay within the decay volume, $F_{decay}$, is then dependent on their lifetimes and boosts.

We calculate $N_{\pi_D}$ as follows. Adopting the simplifying assumption that the $B$ mesons are produced close to threshold, such that their transverse momentum is very small compared to their lab frame longitudinal momentum, we take the $B$ meson momentum distribution from Ref.~\cite{CERN-SHiP-NOTE-2015-009}. For a two-body $B \to K^{(*)} \pi_D$ decay, the magnitude of the dark pion momentum in the rest frame of the $B$ meson is
\begin{equation}
p_{CM} = \frac{1}{2 m_B} \sqrt{\left(m_B^2 - (m_{K^{(*)}}+m_{\pi_D})^2 \right) \left(m_B^2 - (m_{K^{(*)}}-m_{\pi_D})^2 \right)}.
\end{equation}
For each dark pion mass in a sampled range, an expected distribution of longitudinal and transverse momenta in the frame of the $B$ meson is then found by taking a random sample of angles $\theta_{CM}$ from a flat distribution between $-\pi$ and $\pi$, with the magnitude given by $p_{CM}$. 
Upon boosting these according to the $B$ meson momentum spectrum, we find a lab-frame distribution of longitudinal momenta and angles for the dark pions.
%over a distribution of boosts dictated by the $B$ meson momentum distribution, we find a lab frame distribution of longitudinal momenta and angles for the dark pions. 
From these we find $\epsilon_{geom}$ and a distribution of lab-frame decay lengths $L$, for each dark pion mass. The probability of a dark pion decaying within the SHiP decay volume, for a particular value of $L$, is 
\begin{equation}
p_{decay}(L) = \exp\left(-\frac{L_1}{L} \right) - \exp\left(-\frac{L_2}{L} \right),
\end{equation}
with $L_1=$50m, $L_2=$110m. This probability must then be convolved with the $L$ distribution to find the total $F_{decay}$.

\begin{figure}
\begin{center}
\begin{subfigure}[b]{0.5\textwidth}
\includegraphics[width=0.95\textwidth]{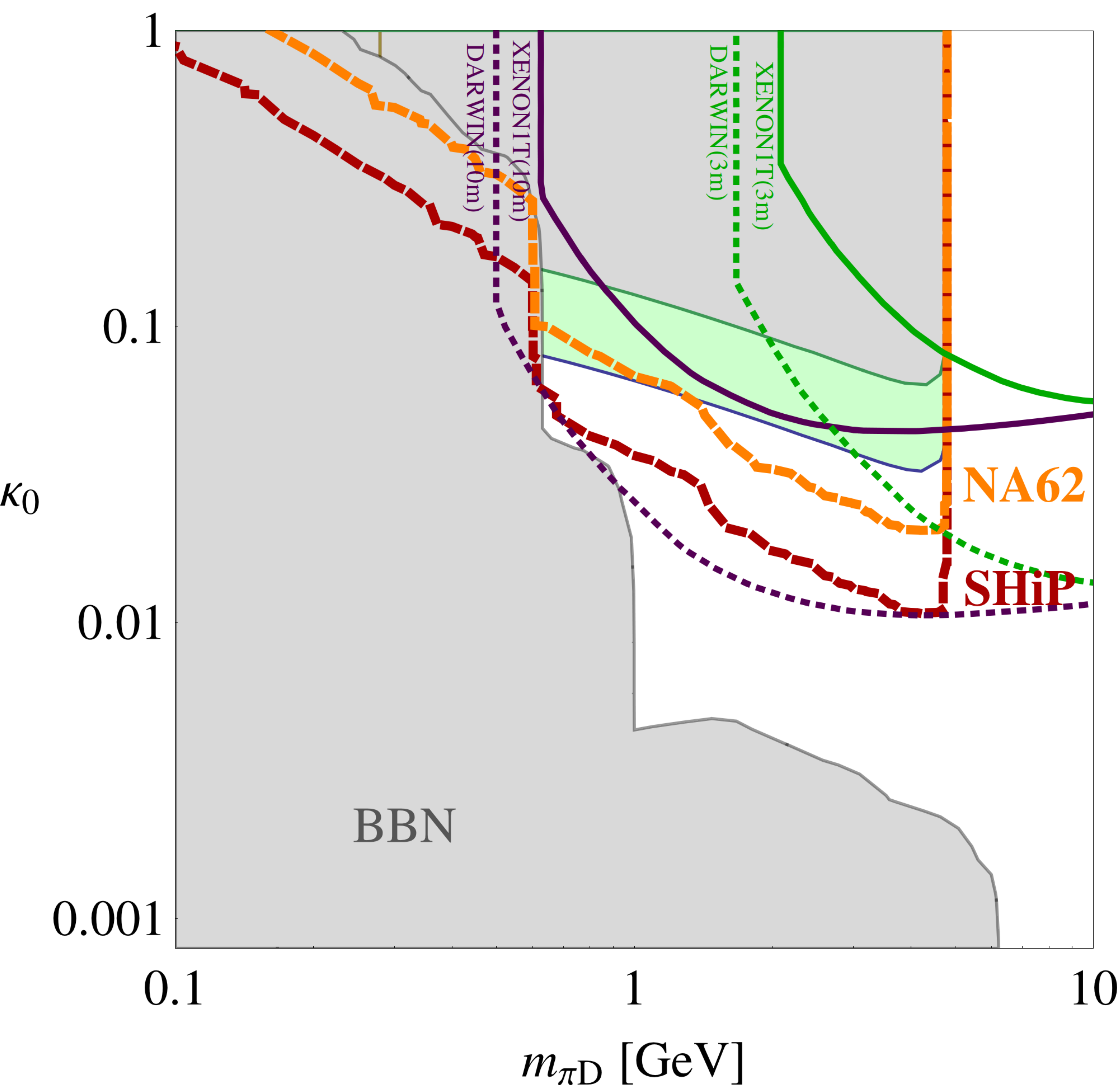}
\caption{``13'' scenario}
\end{subfigure}\begin{subfigure}[b]{0.5\textwidth}
\includegraphics[width=0.95\textwidth]{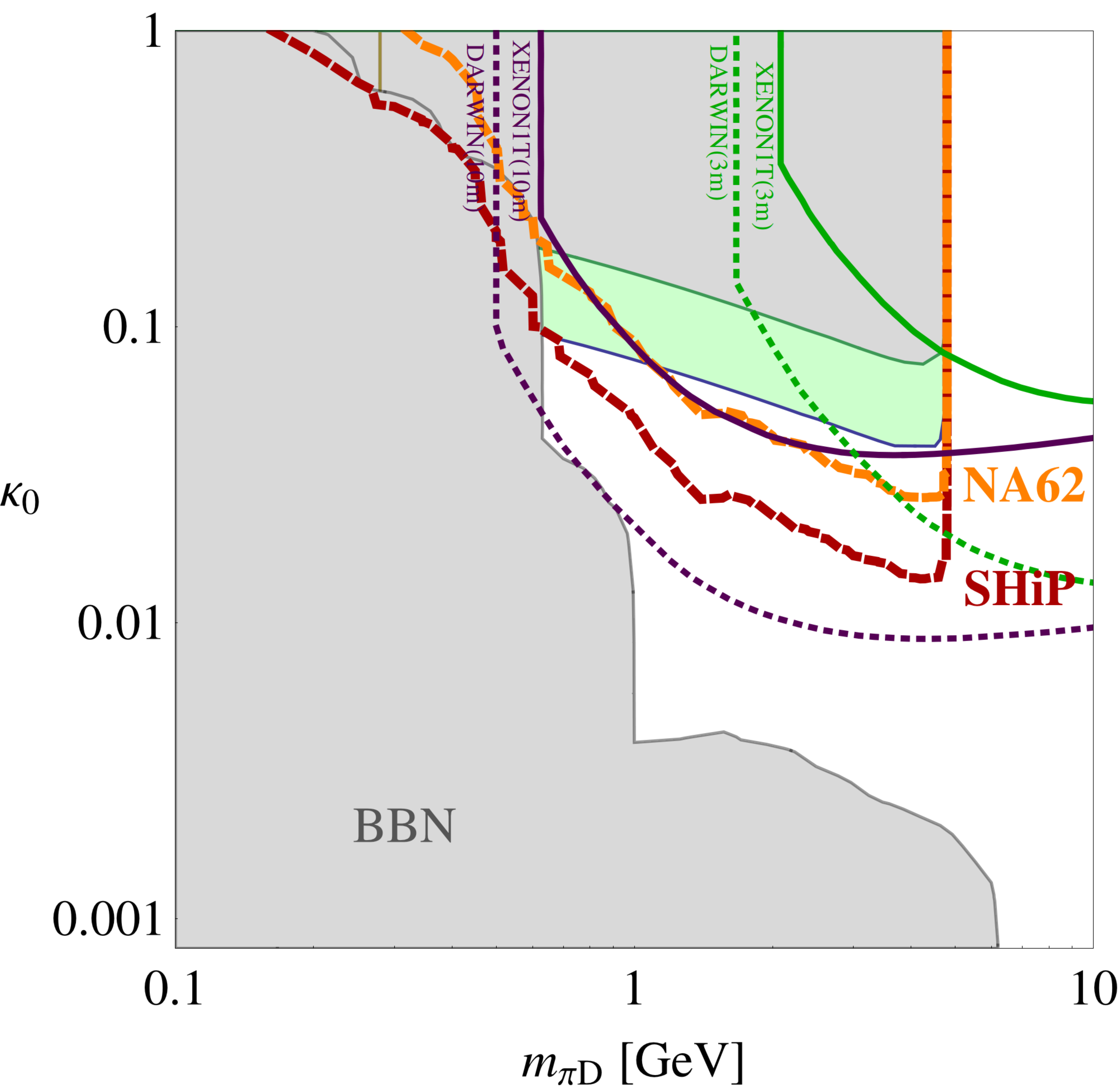}
\caption{``23'' scenario}
\end{subfigure}
\caption{Within the dark red (orange) region, more than 3 dark pion decays are expected over the full run in SHiP (NA62 beam-dump mode). Overlaid are current (solid) and projected (dashed) dark matter direct detection bounds. The bounds in green are calculated under the assumption that the dark protons have a mass of $3 m_{\pi_D}$, while in purple the dark protons are assumed to have a mass of $10 m_{\pi_D}$. The greyed regions are excluded by BBN or flavour constraints. The light green region is where upcoming measurements of ${\rm Br}(B \to K \bar{\nu} \nu)$ at Belle II is expected to have sensitivity. The parameter space in the left (right) plot corresponds to the ``13'' (``23'') scenario, and taking $M_X=1$ TeV, $f_D=m_{\pi_D}$. }
\label{fig:ship}
\end{center}
\end{figure}

The parameter space regions for which more than 3 dark pions decay within the SHiP decay volume are shown in Figure~\ref{fig:ship} by the dark red line. The background expectation over the full run is 0.1 events~\cite{Alekhin:2015byh}, so 3 events corresponds to the expected exclusion region at over $95\%$ confidence level. Of our flavour benchmarks (see Eqns~\eqref{params12}-\eqref{params23} and Table~\ref{tab:decaylengths}), only the ``13'' and ``23'' flavour scenarios are represented here. This is because the flavour structure of the ``aligned'' and ``12'' scenarios prevent the dark pions which are produced in $B$ decays from decaying directly to SM hadrons through the $\kappa$ Yukawa coupling without additional SM flavour-changing interactions, so they will be long-lived. By contrast in the ``13'' and ``23'' scenarios, the flavour misalignment which couples $Q_3$ to $s$ or $d$ (and $b$ to $Q_1$ or $Q_2$) ensures that the dark pions produced in $B$ decays can decay directly back to SM pions and kaons, with decay lengths of the order of fixed-target detector scales. 

The NA62 experiment~\cite{NA62:2017rwk}, designed to measure the very rare decay $K^+\to~\pi^+\bar{\nu} \nu$, can also be operated in ``dump mode'' to search for hidden sector particles~\cite{Dobrich:2017yoq}. To operate in dump mode, the target of NA62 can be lifted so that the 400 GeV proton beam hits the Cu collimator, located 20m downstream. The decay volume is about 80m from the dump, with a length of 65m. The dataset from a few hours long run in dump mode performed in November 2016 provides important information on expected backgrounds, and suggests that an upstream veto in front of the decay volume could reduce backgrounds to close to zero~\cite{NA62beamdump}. The proposed total number of protons on target in beam dump mode is $10^{18}$. In the same way as the SHiP regions, we calculate the number of dark pions expected to decay within the decay volume over the full NA62 dump mode run, in this case taking $\theta_{max}=0.05$. We use the same $B$-meson momentum distribution as for the SHiP case (Ref.~\cite{CERN-SHiP-NOTE-2015-009}), since the proton beam energy is the same.

As noted in the discussions of Sec.~\ref{sec:BandKdecays}, for some regions of parameter space dark pions can be produced in kaon decays and show up in measurements of $K^+ \to \pi^+ \nu \nu$, which will be measured to within $10\%$ at NA62. But these regions are already disfavoured by BBN considerations, making production from $B$ decays a more promising scenario.

For SHiP and NA62 running in dump mode, the decay of $\pi_D$ to any visible final states can be used to probe the model. Instead hidden sector particles can also be searched for in NA62 parasitically during normal operation. In that case there will be a significant background from kaons reaching the decay volume, so searches for new states have to trigger on particular final states. The branching ratios for different dark pion decay modes, calculated using the chiral perturbation theory picture for the final-state mesons, are displayed in Figure~\ref{fig:BRs13} and Figure~\ref{fig:BRs23} in the Appendix, where information about the dark pion species produced in these decays is also given. Decays to $\pi K$ and $3 \pi$ dominate for dark pion masses below $\approx 2$~GeV. This is a direct consequence of the CP-oddness of the dark pion, which forbids decays to CP-even final states such as $\pi^+\pi^-$, and sets it apart from light CP-even scalars or dark photons. Furthermore, different from some axion-like particles (ALPs), decays to di-photon final states are loop suppressed here relative to the leading decay modes~\cite{Dobrich:2015jyk}. Thus dark pions are not only testable at NA62 and other fixed target experiments, they also have a unique signature that sets them apart from other light particle scenarios.

\section{Summary of constraints and conclusions}
\label{sec:conclusions}

Dark pions appear as lightest bound states of confining dark sectors with approximate chiral symmetries, and therefore are essential to understanding the phenomenology of such dark sectors. Here we have for the first time studied scenarios where a flavourful portal to the dark sector imposes a flavour structure on the dark pions, re-analysed cosmological and astrophysical constraints on those scenarios, and their impact on laboratory searches for flavoured dark sectors. 

The main results of our work are summarised in Figure~\ref{fig:ship}. After taking into account constraints from rare meson decays, big bang nucleosynthesis, and dark matter direct detection, a wedge shaped region of parameter space for dark pion masses below 10~GeV remains allowed. A large part of this region will be probed in the future by searches for rare $B$ meson decays at Belle II and in the fixed target experiments NA62 and SHiP. Here in particular the leading discovery channels are $\pi_D \to K^\pm \pi^{\mp}$ and $\pi_D \to \pi^+ \pi^- \pi^0$, which would directly give information about the CP nature of the newly discovered resonance. 

For lower dark pion masses, the upcoming measurement of the $K^+ \to \pi^+ \bar{\nu} \nu$ branching ratio at NA62 will probe very small Yukawa couplings. This region of parameter space however is already in conflict with BBN constraints, i.e.~either the model or its cosmological history would have to be adjusted if evidence for new physics appears in this channel. 

The LHC experiments continue to offer the best opportunities to discover dark pions with masses above the bottom quark threshold. We have shown here that the emerging jet signature can be realised with a realistic flavour structure, but also that the signals can be richer and carry additional information about the underlying model. One characteristic new feature are dark showers which emerge with more than one characteristic time scale, and where the flavour composition of the emerging shower changes with the distance from the interaction point. 

Searches for dark showers can be complemented by searching for decays of individual dark pions decaying in the LHC detectors, in particular at LHCb with its low trigger thresholds and accurate particle reconstruction. Some work in this direction has recently been published~\cite{Gligorov:2017nwh,Pierce:2017taw}, and we expect that similar studies can also constrain our flavoured dark sector scenario. 

\section*{Acknowledgements}

We would like to thank
W.~Bensalem, J.~Martin Camalich, A.~Carmona, C.~Cesarotti, B.~Doebrich, U.~Haisch, J.~Kamenik, M.~Neubert, M.~Selch, B.~Stefanek, D.~Stolarski, and D.~Sutherland for useful discussions and comments on the manuscript. SR would like to thank the CERN theory group and UCSB physics department for hospitality while some of this work was done.
The research reported here has been supported by the Cluster of Excellence Precision Physics, Fundamental Interactions and Structure of Matter (PRISMA -- EXC 1098).

\appendix

\section{Dark chiral perturbation theory and dark pion decays}
\label{appendix:chpt}
In this Appendix the decay rates of dark pions to SM pions and kaons are calculated from the chiral Lagrangian in Equation~\eqref{eqn:chpt}.

\subsection*{Decays to three pions}
If the $\kappa$ matrix is real, decays to two pions are forbidden by $CP$ conservation. So decays to hadrons are only possible if the dark pions have a mass greater than $3 m_{\pi}$, at which point decays to three pions can occur.

At third order in the $\Pi$ fields, the SM current expands to give
\begin{equation}
\left[ U\partial^{\mu}U^{\dagger}\right]_{(3)} = \frac{4 i}{f_{\pi}^3}\left(\Pi \cdot \Pi \cdot d\Pi + d\Pi \cdot \Pi \cdot \Pi - 2 \Pi \cdot d\Pi \cdot \Pi \right),
\end{equation}
where the SM $U$ and $\Pi$ fields are as defined in Eqns.~\eqref{eqn:chiraldefs}.
Thus the Lagrangian terms describing interactions between a dark pion and three mesons are always of the form (with $\pi_D$ representing any dark pion within $\Pi_D$ and $\pi_k$ representing any SM meson within $\Pi$):
\begin{equation}
\label{eqn:3pionamp}
\mathcal{L}\supset \alpha \frac{1}{2M_X^2}\frac{f_D}{f_{\pi}} d\pi_D \left(\pi_1 \pi_2 d\pi_3 + d\pi_1 \pi_2 \pi_3 -2 \pi_1 d\pi_2 \pi_3\right),
\end{equation}
where $\alpha$ is some constant factor (found on performing the expansion) which depends on the identity of the dark pions and mesons in question, and on the $\kappa$ matrices. This leads to a decay amplitude
\begin{equation}
A_{\pi \pi \pi}= \alpha \frac{1}{2M_X^2}\frac{f_D}{f_{\pi}} \left(p_D \cdot p_3 +p_D \cdot p_1 -2 p_D \cdot p_2 \right),
\end{equation}
with $p_D$ the momentum of the dark pion and $p_k$ the momenta of the three final state mesons respectively. For three mesons with the same mass $m_{\pi}$, this can be written
\begin{equation}
A_{\pi \pi \pi}= \alpha \frac{f_D}{2f_{\pi} M_X^2} \, \frac{1}{2}\left(-m_{\pi_D}^2-3 m_{\pi}^2 + 3 s_{13} \right),
\end{equation}
where $s_{ij} \equiv (p_i + p_j)^2$.
The decay rate to three pions is then~\cite{RevModPhys.36.595}
\begin{align}
\label{rate}
\Gamma_{\pi \pi \pi} &= \frac{1}{S}\frac{1}{(2\pi)^3} \frac{1}{32m_{\pi_D}^3}\int|A_{\pi \pi \pi}|^2 \delta \left(s_{12}+s_{13}+s_{23}-(m_{\pi_D}^2+3m_{\pi}^2)\right)
\Theta(\Delta_3) \, ds_{12}ds_{23}ds_{13}
\end{align}
where 
\begin{equation}
\Delta_3 \equiv \mathrm{det}\begin{pmatrix}
m_{\pi}^2 & p_{12} & p_{13} \\ 
p_{21} & m_{\pi}^2 & p_{23} \\
p_{31} & p_{32} & m_{\pi}^2
\end{pmatrix},
\end{equation}
with $p_{ij} \equiv p_i\cdot p_j$. The factor $\Theta(\Delta_3)$ defines the physical region. Using the definition of $\Delta_3$ we can find the allowed values of (e.g.)~$s_{23}$ for a given value of (e.g.)~$s_{13}$, and then integrate over $s_{13}$. In this way the rate becomes
\begin{align}
\Gamma_{\pi \pi \pi} &= \frac{1}{S}\frac{1}{(2\pi)^3} \frac{1}{32m_{\pi_D}^3}\int_{4m_{\pi}^2}^{(m_{\pi_D}-m_{\pi})^2} \mu_{\pi\pi\pi}|A_{\pi\pi\pi}|^2 ds_{13} \\
&=\frac{1}{(2\pi)^3} \frac{1}{32m_{\pi_D}^3}\int_{4m_{\pi}^2}^{(m_{\pi_D}-m_{\pi})^2}\mu_{\pi\pi\pi}\left| \alpha \frac{1}{2M_X^2} \frac{f_D}{f_{\pi}} \, \frac{1}{2}\left(-m_{\pi_D}^2-3 m_{\pi}^2 + 3 s_{13} \right)\right|^2 ds_{13} \\
&=\frac{1}{S}\frac{|\alpha|^2}{4096 \pi^3 m_{\pi_D}^3 M_{X}^4}\frac{f_D^2}{f_{\pi}^2}\big[ (m_{\pi_D}^2+3m_{\pi}^2)^2 I_0 - 24 m_{\pi}^2 (m_{\pi_D}^2+3m_{\pi}^2)I_1 +144 m_{\pi}^4 I_2 \big] \nonumber\\
&  ~~\times\sqrt{\frac{(m_{\pi_D}^2-9m_{\pi}^2)(m_{\pi_D}^2-m_{\pi}^2)(m_{\pi_D}-3m_{\pi})^3(m_{\pi_D}+m_{\pi})^3}{4m_{\pi}^2}}
\end{align}
where $S$ is a symmetry factor
\begin{equation}
S=\begin{cases}
6 &\text{for~} \pi_D \to \pi^0 \pi^0 \pi^0\\
1 &\text{for~} \pi_D \to \pi^0 \pi^+ \pi^-
\end{cases}
\end{equation}
and defining
\begin{equation}
\mu_{\pi\pi\pi}=\sqrt{\frac{(s_{13}-4m_{\pi}^2)(s_{13}-(m_{\pi_D}+m_{\pi})^2)(s_{13}-(m_{\pi_D}-m_{\pi})^2)}{s_{13}}}\,,
\end{equation}
\begin{align}
I_n &\equiv B\left(\frac{3}{2},\frac{3}{2}\right) F_1 \left(\frac{3}{2},\frac{1}{2}-n,-\frac{1}{2},3,\frac{(3m_{\pi}-m_{\pi_D})(m_{\pi_D}+m_{\pi})}{4m_{\pi}^2}, \frac{(3m_{\pi}-m_{\pi_D})(m_{\pi_D}+m_{\pi})}{(m_{\pi_D}+3m_{\pi})(m_{\pi_D}-m_{\pi})}\right)\,,\nonumber
\end{align}
where $B(x,y)$ is the Euler Beta function, and $F_1$ is the Appell hypergeometric function~\cite{gradshteyn2007}.\footnote{Both $F_1$ and $B(x,y)$ are built-in to Mathematica, taking arguments in the order given here.} The above calculation can be straightforwardly applied also to decays to three kaons.

\subsection*{Decays to two mesons of different mass}
Once kaons are kinematically available, decays $\pi_D \to K \pi$ become possible. Expanding the SM current to second order in the $\Pi$ fields, the Lagrangian terms for these interactions are of the form

\begin{equation}
\mathcal{L} \supset \beta \frac{if_D}{2 M_X^2}(\partial_{\mu} \pi_D) \left(K (\partial^{\mu} \pi)- (\partial^{\mu} K)\pi \right)
\end{equation}
where $\beta$ is some constant factor which depends on the identity of the dark pions and mesons in question, and on the $\kappa$ matrices. This leads to a decay amplitude
\begin{align}
A_{K\pi}&= \beta \frac{if_D}{2 M_X^2} \left(p_D \cdot p_1 - p_D \cdot p_2 \right)\\
&= \beta \frac{if_D}{2 M_X^2} \left(m_K^2-m_{\pi}^2\right).
\end{align}
The decay rate is
\begin{align}
\Gamma_{K \pi} &= \frac{1}{8\pi} \frac{|p|}{m_{\pi_D}^2} |A_{K\pi}|^2 \\
&= |\beta|^2 \frac{f_D^2}{64\pi^2 M_X^4} \frac{\left(m_K^2-m_{\pi}^2\right)^2}{m_{\pi_D}^3}\sqrt{\left( m_{\pi_D}^2-(m_K+m_{\pi})^2\right) \left( m_{\pi_D}^2-(m_K-m_{\pi})^2\right)}.
\end{align}

\subsection*{Decays to two pions and a kaon}
In some regions of parameter space, decays to three non-identical mesons are important. The outline of the calculation for decays to two pions and a kaon follows, but it can be straightforwardly generalised to any three-body decay in which two of the mesons have the same mass. 

The amplitude~\eqref{eqn:3pionamp} applies in this case, and assigning $\pi_1$ and $\pi_3$ to be pions and $\pi_2$ to be a kaon, this becomes
\begin{equation}
A_{K\pi\pi}= \alpha \frac{1}{2M_X^2}\frac{f_D}{f_{\pi}} \left(-m_K^2+s_{13} \right),
\end{equation}
where as before $s_{ij} \equiv (p_i + p_j)^2$, and $\alpha$ is a constant which depends on the identity of the dark pions and mesons in question, and on the $\kappa$ matrices. This leads to the decay rate
\begin{align}
\Gamma_{K\pi\pi} = \frac{1}{S}\frac{1}{(2\pi)^3} \frac{1}{32m_{\pi_D}^3}\int & |A_{K\pi\pi}|^2 \delta \left(s_{12}+s_{13}+s_{23}-(m_{\pi_D}^2+2m_{\pi}^2+m_K^2)\right) \nonumber \\
& \times \Theta(\Delta_3) \, ds_{12}ds_{23}ds_{13}
\end{align}
where $S=2$ if there are two identical particles (i.e.~$\pi^0 \pi^0$) in the final state, $S=1$ otherwise. Now
\begin{equation}
\Delta_3 \equiv \mathrm{det}\begin{pmatrix}
m_{\pi}^2 & p_{12} & p_{13} \\ 
p_{21} & m_{K}^2 & p_{23} \\
p_{31} & p_{32} & m_{\pi}^2
\end{pmatrix}
\end{equation}
with $p_{ij} \equiv p_i\cdot p_j$. All the integration limits can be set to $[4m_{\pi}^2,(m_{\pi_D}-m_{\pi})^2]$; the factor $\Theta(\Delta_3)$ defines the physical region. To our knowledge this integral has no analytic solution, and must be evaluated numerically.

\subsection*{Kinematic regions}
\begin{figure}
\begin{subfigure}[b]{0.5\textwidth}
\includegraphics[width=0.9\textwidth]{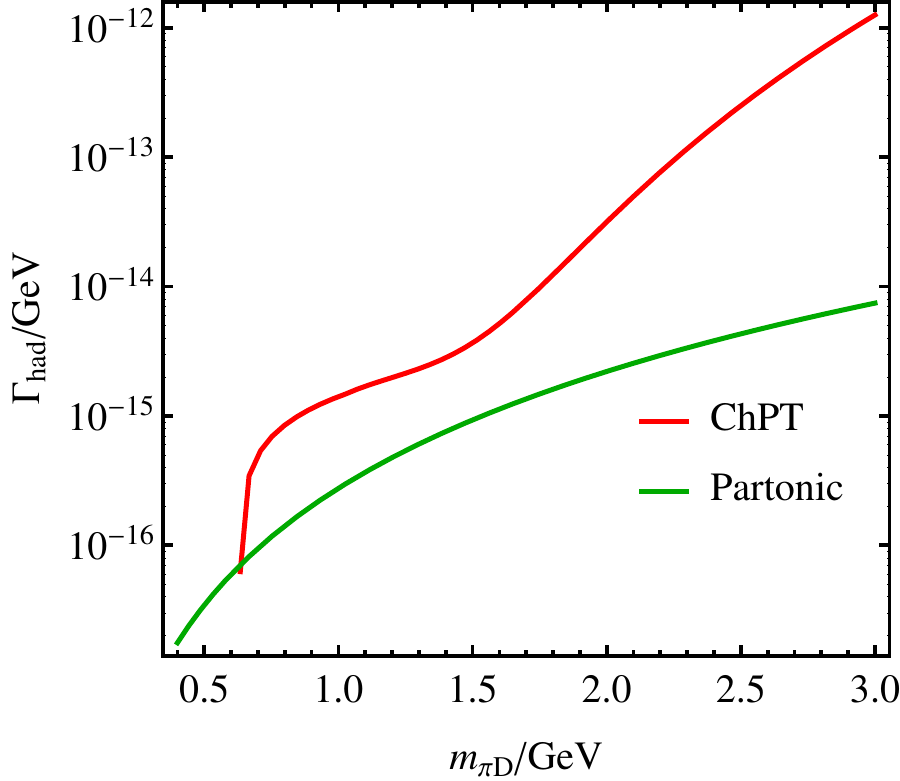}
\caption{$\pi_{D_1}$ and $\pi_{D_2}$}
\end{subfigure}
\begin{subfigure}[b]{0.5\textwidth}
\includegraphics[width=0.9\textwidth]{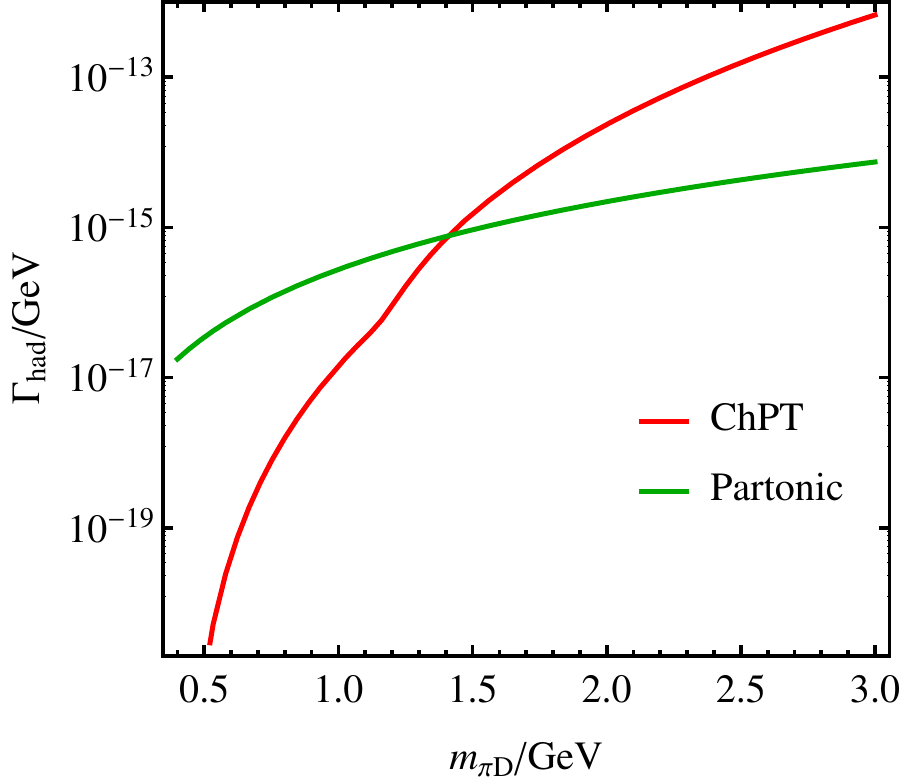}
\caption{$\pi_{D_3}$ and $\pi_{D_8}$}
\end{subfigure}
\caption{Plots showing the width to hadrons calculated in both the chiral and partonic picture, for $\kappa=\mathbb{1}_{3\times 3}$, $f_D=m_{\pi_D}$. Plots are shown for each pion that can decay hadronically with these parameter choices.}
\label{fig:matchingplots}
\end{figure}

We briefly justify choices we make in Sec.~\ref{sec:model} to match a chiral perturbation theory picture and a partonic picture for SM hadrons.
Figure~\ref{fig:matchingplots} shows the total width of the four dark pions that can decay hadronically with the coupling choice $\kappa=\mathbb{1}_{3\times 3}$ (without the need for extra loops or SM flavour violation). The width is calculated using both the chiral perturbation theory Lagrangian~\eqref{eqn:chpt} (red line), and the partonic Lagrangian~\eqref{eqn:partoniclagrangian} (green line). These plots were made assuming $m_X=1$ TeV and $f_D=m_{\pi_D}$. It can be seen that the widths in the two theories agree to within $O(1)$ factors at $m_{\pi_D}\approx 1.5$ GeV.

\subsection*{Branching ratios}

\begin{figure}
\begin{center}
\includegraphics[width=0.5\textwidth]{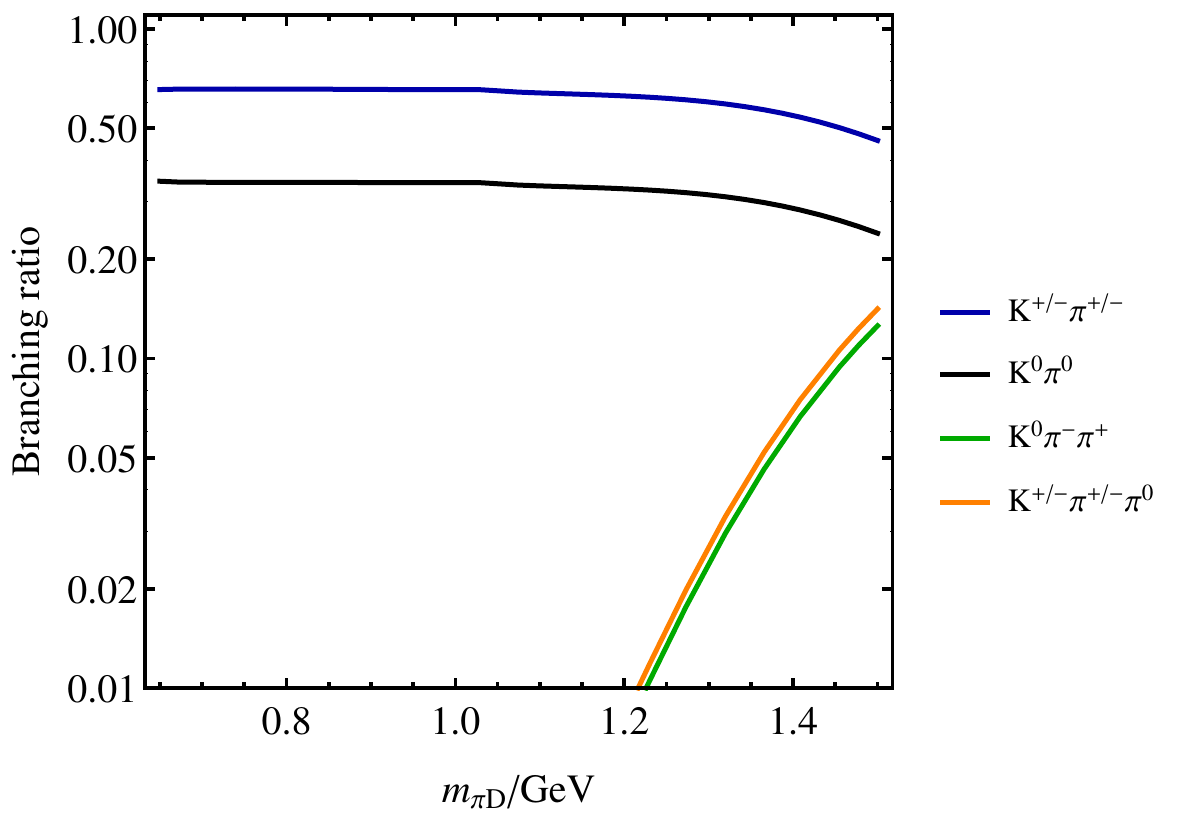}
\caption{Branching ratios for $\pi_{D_1}$, $\pi_{D_2}$, $\pi_{D_6}$ and $\pi_{D_7}$, within the ``13'' scenario.}
\label{fig:BRs13}
\end{center}
\end{figure}

The branching ratios for the dark pions produced in $B \to K \pi_D$ decays within the ``13'' scenario are shown in Figure~\ref{fig:BRs13}. For this particular flavour scenario, the dark pions $\pi_{D_6}$ and $\pi_{D_7}$ are produced with $99.75\%$ probability in these decays, while $\pi_{D_1}$ and $\pi_{D_2}$ are produced with $0.25\%$ probability.

\begin{figure}
\begin{subfigure}[b]{0.5\textwidth}
\includegraphics[height=5.2cm]{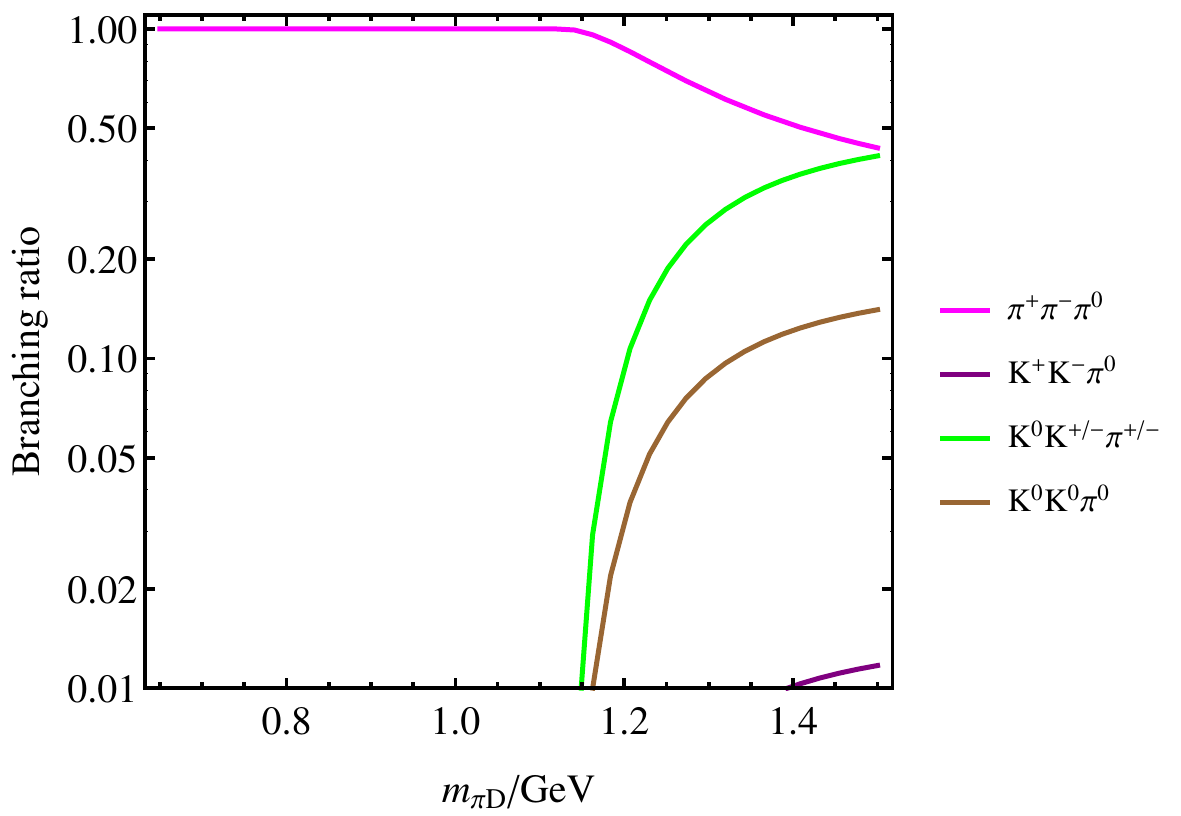}
\caption{$\pi_{D_3}$ and $\pi_{D_8}$}
\end{subfigure}\begin{subfigure}[b]{0.5\textwidth}
\includegraphics[height=5.2cm]{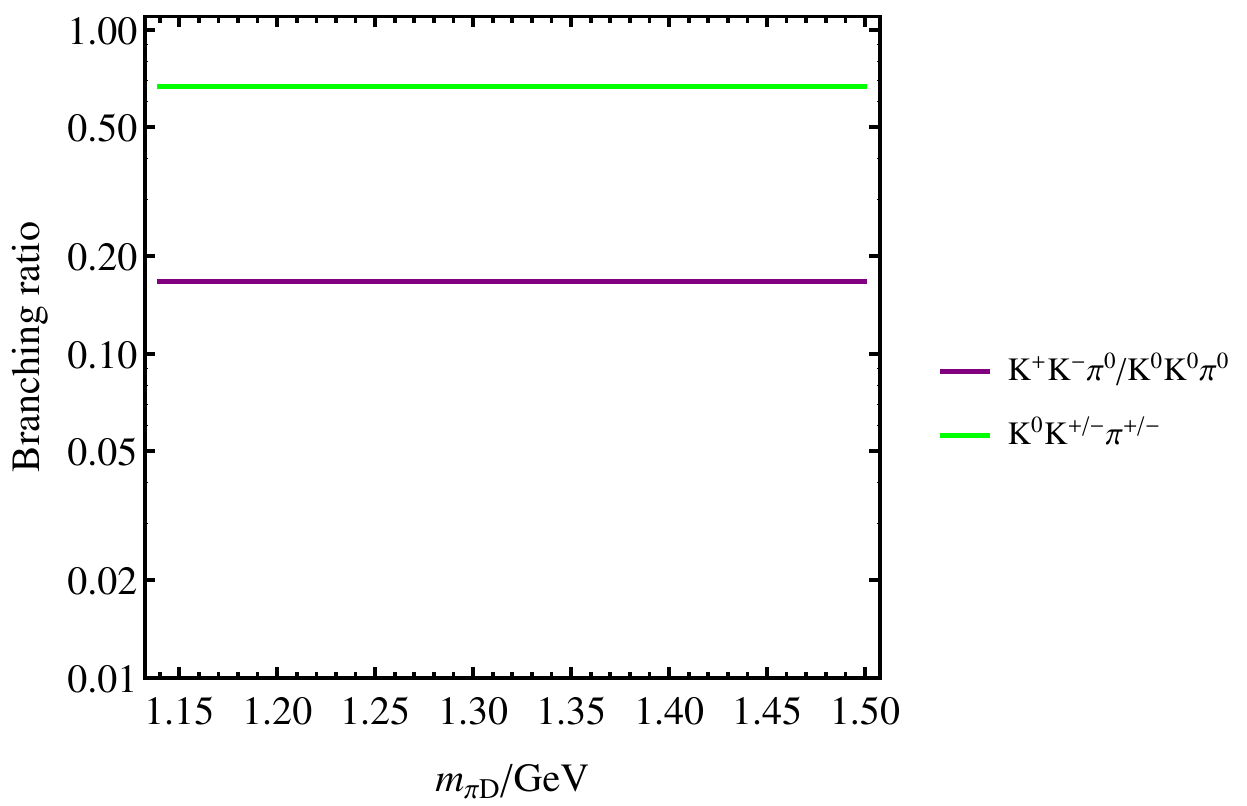}
\caption{$\pi_{D_6}$ and $\pi_{D_7}$}
\end{subfigure}
\caption{Branching ratios for (left) $\pi_{D_3}$ and $\pi_{D_8}$, and (right) $\pi_{D_6}$ and $\pi_{D_7}$, within the ``23'' scenario.}
\label{fig:BRs23}
\end{figure}

Figure~\ref{fig:BRs23} displays branching ratios for the dark pions produced in $B \to K \pi_D$ decays within the ``23'' scenario. For this  flavour scenario, the dark pions $\pi_{D_6}$ and $\pi_{D_7}$ are produced with $91.3\%$ probability in these decays, while $\pi_{D_3}$ and $\pi_{D_8}$ are produced with $8.67\%$ probability.

\section{Range of lifetimes}
\begin{figure}
\begin{center}
\includegraphics[width=.32\textwidth]{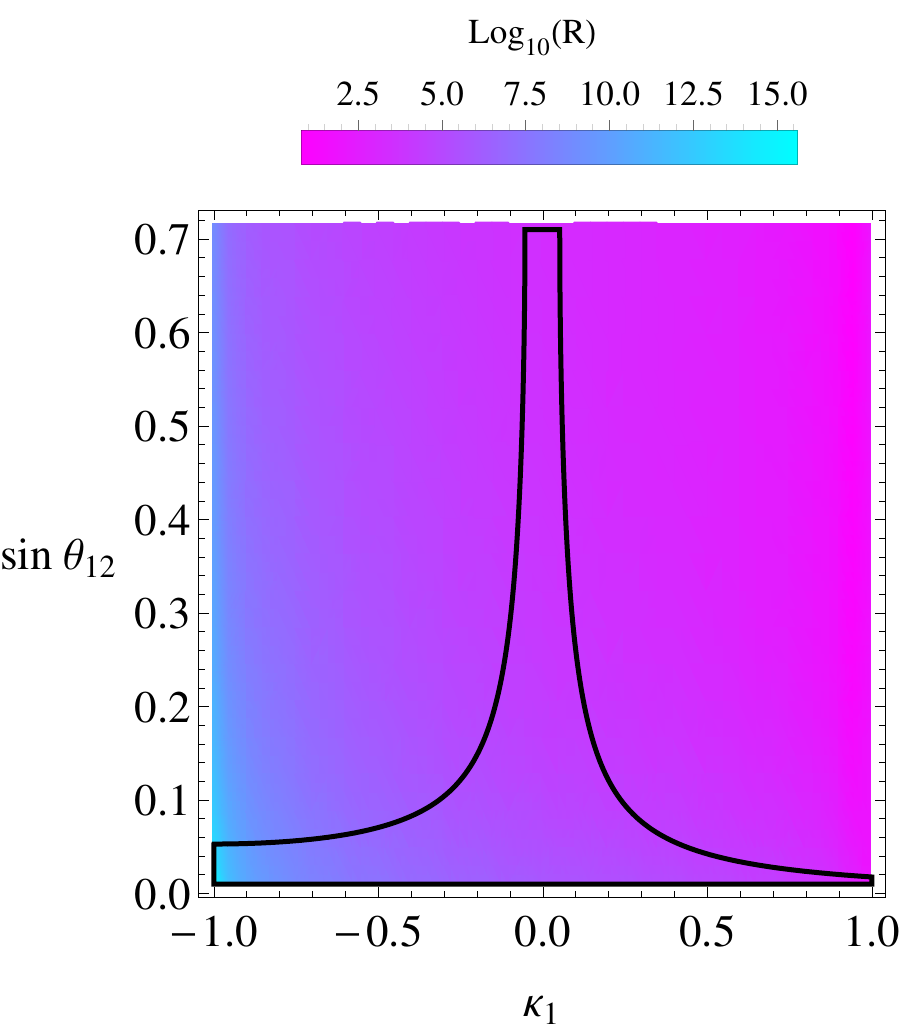} 
\includegraphics[width=.32\textwidth]{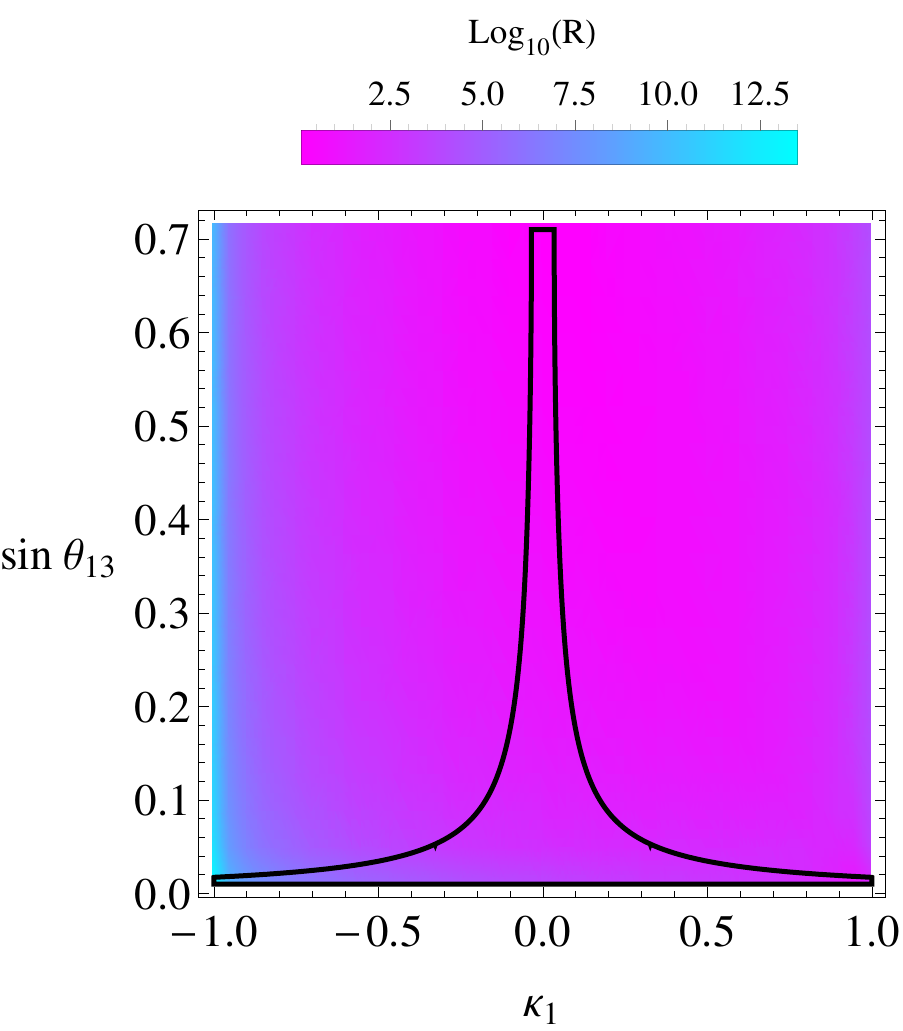} 
\includegraphics[width=.32\textwidth]{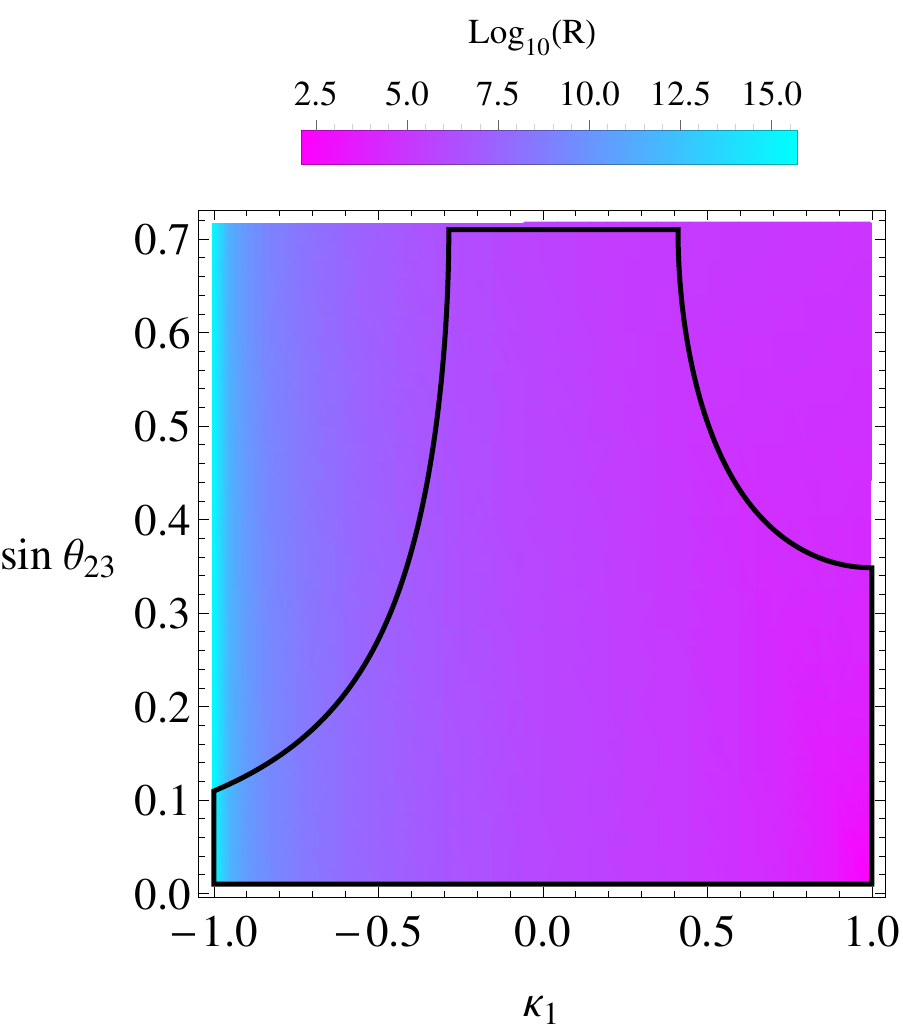}
\caption{
Correlation of dark pion lifetimes and departure from the flavour aligned limit. The ratio between the longest and shortest dark pion proper lifetime is denoted $R$ and shown by the heat map, for each of the scenarios defined in Eqns.\eqref{params12} - \eqref{params23}. The regions enclosed by the black curves are allowed by meson mixing constraints. The dark pion and mediator masses are set to $m_{\pi_D}=10$ GeV, $m_X=1$ TeV, such that constraints from $\Delta F = 1$ decays are absent. 
}
\label{fig:lifetimeratios}
\end{center}
\end{figure}

It is clear that in the aligned limit, a strong hierarchy of dark pion lifetimes appears. Deviations from the aligned case are subject to flavour mixing constraints, but still allow for a more homogeneous range of lifetimes, as shown in Figure~\ref{fig:lifetimeratios}.

\section{Meson mixing parameter analysis}
Here we detail proof of the statements made in Sec.~\ref{sec:mesonmixing} about scenarios for the parameters of the coupling matrix that do not produce contributions to meson mixing. The contributions to $K^0-\bar{K}^0$ mixing are proportional to
\begin{equation}
\left(\sum_{\alpha=1}^3  \kappa_{\alpha s}\kappa^*_{\alpha d} \right)^2,
\end{equation}
with corresponding expressions for $B_{(d,s)}$ mixing (with $s \to b$, $d \to (d,s)$). 
For a scenario in which $D\propto \mathbb{1}_{3\times 3}$ ($\kappa_1=\kappa_2=0$), this gives:
\begin{align*}
\left(\sum_{\alpha=1}^3  \kappa_{\alpha q}\kappa_{\alpha q^{\prime}}^* \right)^2&=\left(\sum_{\alpha =1}^3 [DU]^{\dagger}_{q^{\prime}\alpha}[DU]_{\alpha q} \right)^2 \\
&=\left(\sum_{\alpha,i,j=1}^3 U^{\dagger}_{q^{\prime}i}D^{\dagger}_{i\alpha }D_{\alpha j}U_{j q} \right)^2 \\
&=\kappa_0^4\left(\sum_{\alpha=1}^3 U^{\dagger}_{q^{\prime}i}U_{i q} \right)^2 \\
&=0\,.
\end{align*}
Therefore there is no meson mixing in this case, irrespective of the values of the angles $\theta_{ij}$. 
If $\kappa_1=\kappa_2$, such that $\Delta_{12}=0$, but $\Delta_{13}$ and $\Delta_{23}$ are non-zero:
\begin{align*}
\allowdisplaybreaks
\left( \sum_{\alpha =1}^3 \kappa_{\alpha q}\kappa_{\alpha q^{\prime}}^* \right)^2&=\left([DU]^{\dagger}_{q^{\prime}\alpha}[DU]_{\alpha q}  \right)^2 \\
&=\left(U^{\dagger}_{q^{\prime}i}D^{\dagger}_{i\alpha }D_{\alpha j}U_{j q} \right)^2 \\
&=\left((\kappa_0+\kappa_1)^2 U^{\dagger}_{q^{\prime}1}U_{1q} +(\kappa_0+\kappa_2)^2 U^{\dagger}_{2 q^{\prime}}U_{q2} +(\kappa_0-\kappa_1-\kappa^2)^2 U^{\dagger}_{q^{\prime}3}U_{3 q} \right)^2 \\
&=\left((\kappa_0+\kappa_1)^2 U^{\dagger}_{q^{\prime}1}U_{1 q} +(\kappa_0+\kappa_1)^2 U^{\dagger}_{q^{\prime}2}U_{2 q} +(\kappa_0-2\kappa_1)^2 U^{\dagger}_{q^{\prime}3}U_{3 q} \right)^2 \\
&=\left((\kappa_0+\kappa_1)^2 (U^{\dagger}_{q^{\prime}1}U_{1 q} + U^{\dagger}_{q^{\prime}2}U_{2 q}) +(\kappa_0-2\kappa_1)^2 U^{\dagger}_{q^{\prime}3}U_{3 q} \right)^2 \\
&=\Bigg((\kappa_0+\kappa_1)^2 \left([U^{\dagger}_{23}]_{q^{\prime}i}[U^{\dagger}_{13}]_{ij}([U^{\dagger}_{12}]_{j 1}[U_{12}]_{1k}+[U^{\dagger}_{12}]_{j 2}[U_{12}]_{2k}) [U_{13}]_{kl}[U_{23}]_{l q^{\prime}}]\right)\\
&+(\kappa_0-2\kappa_1)^2 \left([U^{\dagger}_{23}]_{q^{\prime}i}[U^{\dagger}_{13}]_{ij}[U^{\dagger}_{12}]_{j 3}[U_{12}]_{3k} [U_{13}]_{kl}[U_{23}]_{l q}]\right) \Bigg)^2 \\
&=\Bigg((\kappa_0+\kappa_1)^2 \left([U_{23}^{\dagger}]_{q^{\prime}i}[U_{13}^{\dagger}]_{ij}(\delta_{j 1}\delta_{k 1}+\delta_{j 2}\delta_{k 2}) [U_{13}]_{kl}[U_{23}]_{l q}]\right)\\
&+(\kappa_0-2\kappa_1)^2 \left([U_{23}^{\dagger}]_{q^{\prime}i}[U_{13}^{\dagger}]_{i3} [U_{13}]_{3 l}[U_{23}]_{l q}]\right) \Bigg)^2,
\end{align*}
where repeated indices are summed over. The mixing matrix $U_{12}$ has dropped out, meaning that  there is no dependence on $s_{12}$ or $c_{12}$, and to prevent any contribution to meson mixing in this scenario we need only $s_{13, 23}=0$. Similar arguments follow for cases in which $\Delta_{13}$ or $\Delta_{23}$ are zero.

\bibliographystyle{JHEP}
\bibliography{DMflavour}
\end{document}